# The Virtual Reality Conjecture[1]

**Brian Whitworth**

*Institute of Information and Mathematical Sciences,*
*Massey University, Albany, Auckland, New Zealand*

## Abstract

*We take our world to be an objective reality, but is it? The assumption that the physical world exists in and of itself has struggled to assimilate the findings of modern physics for some time now. For example, an objective space and time would just "be", but in relativity, space contracts and time dilates. Likewise objective "things" should just inherently exist, but the entities of quantum theory are probability of existence smears, that spread, tunnel, superpose and entangle in physically impossible ways. Cosmology even tells us that our entire physical universe just "popped up", from nowhere, about 14 billion years ago. This is not how an objectively real world should behave! Yet traditional alternatives don't work much better. That the world is just an illusion of the mind doesn't explain its consistent realism and Descartes dualism, that another reality beyond the physical exists, just doubles the existential problem. It is time to consider an option we might normally dismiss out of hand. This essay explores the virtual reality conjecture, that the physical world is the digital output of non-physical quantum processing. It finds it neither illogical, nor unscientific, nor incompatible with current physics. In this model, quantum entities are programs, movement is the transfer of processing, interactions are processing overloads and the fields of physics are network properties. It has no empty space, no singularities and all the conservations of physics just conserve processing. Its prediction, that the collision of high frequency light in a vacuum can create permanent matter, will test it. If the physical world has the properties of a processing output, physics must rewrite the story behind its equations.*

## Introduction

Computers today simulate entire worlds, with their own time, space and objects, but that our 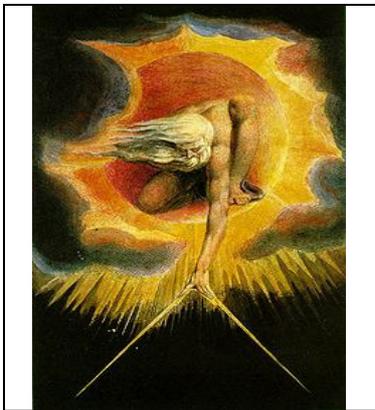 world is so is normally a topic of science fiction, not physics. Yet the idea that the world is illusory has a long history. In Buddhism the world expresses the Universal Mind, in Hinduism it is Maya or God's play, and to Plato it was just shadows flickering on a wall[1]. That the world is digital is also not new, as Pythagoras saw in numbers the non-material essence of the world, Plato felt that "*God geometrizes*" and Gauss that "*God computes*", as Blake's "Ancient of Days" measures the world with a compass (Figure 1). The tradition continues today, as Zuse suggests that "space calculates" [5] and others that reality computes[2]. This essay explores the *virtual reality (VR) conjecture*, that the physical world is the digital output of non-physical quantum processing.

*Figure 1. God computes?*

One can contrast Platonic *idealism*, that the seen world reflects a greater unseen one, with Aristotelian *physicalism* that we see what is actually there. Logically, one of these world views must be wrong but after centuries of confrontation, science and religion formed the compromise of *dualism,* that mind and body realms *both* exist, dividing scientists into atheists who believed *only* in the physical world, theists who *also* believed in a non-physical spiritual reality, and agnostics who didn't know.

Today, dualism seems increasingly a theory kludge[2], a marriage of convenience not truth. If different mind and body realms exist but don't interact, what relevance are they to each other? Or if they interact, which came first? If a conscious mind "emerges" from the physical brain, isn't it superfluous? Dualism is currently in retreat before the simpler, non-dual view *that there is only one real world*. Scientists observing this ideological war generally feel that if there is only one world, let it be the one science studies, i.e. the physical.

If everything is physical, could a self-contained physical universe compute itself [3]? A closed system can no more output itself *entirely* than a physical computer can print itself out[4] [1 p6]. If quantum processing generates the physical world, the latter cannot be complete in itself [5]. This contradicts the "prime axiom" of physics, that:

> *There is nothing outside the physical universe* [8].

The VR conjecture implies its antithesis, that:

> *Nothing in the physical universe exists objectively, i.e. of or by itself.*

The hypothesis is that the physical world is not an objective reality but a processing output, where:

1. *Processing:* is the transforming of information[6] [11], i.e. making a new choice.

2. *The physical world:* is empirical reality, as experienced not just by us but by everything.

3. *An objective reality:* is one that self-exists, needing nothing outside itself to exist.

This doesn't, as some naively think, imply physical "hardware" in a metaphysical realm. The quantum world is unlike the physical world we know: quantum states disappear at will so are not constant like matter-energy, entangled entities ignore the speed of light limit of physical reality and superposed states overlap simultaneously in physically contradictory ways. The quantum world is in every way *physically impossible*, so physicality cannot be its reality base. What that basis is however is not this theory's concern. It is about *the world we experience*. That this world is a virtual reality vs. that it is an objective reality are mutually exclusive hypotheses about it that science can evaluate. If science finds that it *cannot* be objectively real, *it must explore if it is virtual*.

## The evidence

Computing tells us how virtual worlds look, so how does our physical world compare? The findings of modern physics suggest a match:

1. *The big bang.* That our universe arose from "nothing" in a time-zero event makes no sense for an objective reality, but every virtual reality boots up from nothing in itself.

2. *The speed of light.* An objective reality has no reason for a maximum speed, but every simulation has a maximum screen refresh rate that limits local transfers.

3. *Planck limits.* An objective space has no reason to be discrete, as our world seems to be at the Planck level, but a virtual space must be so.

4. *Non-locality.* Effects that *instantly* affect entities *anywhere* in the universe, like entanglement and quantum collapse, are impossible in an objective reality, but a program can alter pixels anywhere on a screen, even one as big as our universe.

5. *Malleable space-time*. Mass and movement shouldn't alter time or space in an objective reality, but in a virtual reality, the processing load of a massive body could dilate virtual time and curve virtual space, given that time and space also arise from processing.

6. *Randomness.* If every physical event is predicted by others, a random quantum event is an impossible "uncaused cause", but a virtual world processor can easily generate random effects.

7. *Empty space is not empty*. In an objective reality, empty space is just "nothing" while our space spawns the virtual particles of the Casimir effect, but space as null processing could do this.

8. *Superposition.* Objective entities cannot spin in two directions at once, as quantum entities do, but a program can instantiate twice to do this.

---

[2] A kludge is a computer system made up of poorly matched components.

9. *Equivalence*. That every electron in our world is *exactly* like every other[7] is unexpected for an objective world, but every electron could be a program from the same code template.

10. *Quantum tunnelling*. An electron "tunnelling" through an impenetrable field barrier is impossible for objects that continuously self-exist, like a coin popping out of a perfectly sealed glass bottle, but a program entity distributed across many "instances" can restart at any one, i.e. "teleport".

If the world acts like a virtual reality not an objective reality, then the duck principle applies:

*If it looks like a duck, and quacks like a duck, then it is a duck.*

## The theory

Current theories of reality [9] include:

a) *Physicalism*. Only the physical world really exists.

b) *Solipsism*[8]. The physical world is a dream or illusion of the mind.

c) *Dualism*. The physical world exists, but there is higher reality beyond it.

Yet logically another non-dual option remains, namely *virtualism*, that a *mind-independent, non-physical reality* outputs the physical world by processing. In this admittedly radical view, the "ghostly" world of quantum theory is real and the physical world is like a screen image thrown up.

Do we then live in the Matrix? In that movie, people knew their reality by the information they received from it, as we know ours. Only when he takes a pill, does the hero fall back into a containing, but also physical, reality to see machines farming human brains in vats. The VR conjecture is *not* that another physical world creates the world we see, but the *opposite*, that physicality itself is virtual. Equally, it is not *solipsism*, that the physical world is just a dream, which Dr Johnson refuted by stubbing his toe on a stone, saying "*I disprove it thus*". Again this theory is the *opposite*, as a quantum reality is indeed "out there" apart from us - it just has no physical properties.

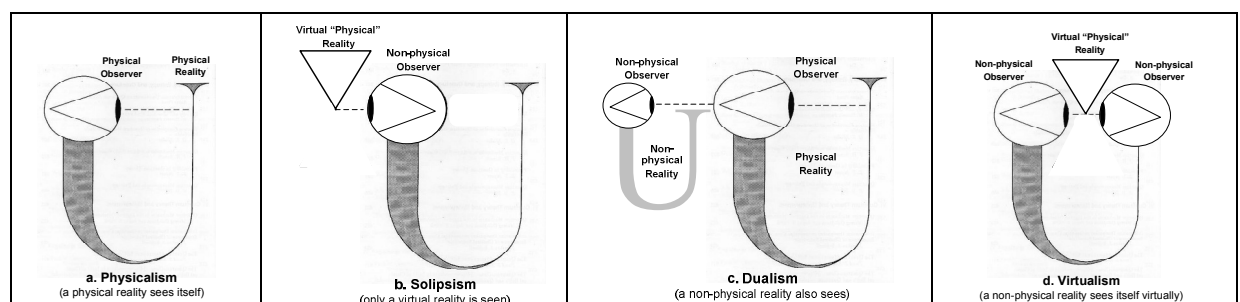

*Figure 2. Universal (U) reality models*

Figure 2 shows the options. In *physicalism* (2a), part of a self-existent physicality sees the rest of it when consciousness mystically "emerges" from physical complexity [10]. In *solipsism* (2b), a self-existent consciousness imagines a dream world, that doesn't really exist at all. In *dualism* (2c), a self-existent physical world exists alongside another also self-existent "higher" reality, also observing it. In *virtualism* (2d), a non-physical quantum reality observes itself using a virtual "physical reality" that has no inherent existence. Yet it is no unreal dream universe with us its existential centre, as *everything* is "knowing" everything else[9], like the most massive of online multiplayer games. The physical world, like a computer game, is a *local reality* - real internally but unreal externally, so stubbing virtual toes can still give virtual pain. In a computer game, the virtual *pixel substrate* does not contain the player, who exists outside of it. This allows for an *observer substrate*[10] as the ultimate existential context, i.e. consciousness could be what allows observation to occur. However this theory doesn't speculate on that, on what is behind the pixels. That a processing system can simulate itself to itself is subtle[11] but not impossible. From our point of view, it is that there is indeed a real world around us - it just isn't the physical world we see.

## The model

If the physical world is a processing output, it should be possible to *reverse engineer* it, i.e. to derive the physics of time, space, light, energy and matter from processing principles. A simulated world needs a screen to produce it, which could be what Wilczek calls:

"*the Grid, that ur-stuff that underlies physical reality*" [12 p111].

The grid is envisioned as a network of dynamically connected nodes, like a cell-phone grid but passing processing dynamic instructions not static data. A photon is a *program entity* that distributes its processing in packets by the computing method of "instantiation"[12]. These *packets* pass between grid nodes to set quantum state *pixels*. If each node *first* gives its processing to *all* its neighbours, grid processing can spread like a wave ripples on a 3D pond, i.e. quantum waves could be processing waves. Not only light, energy and matter but also space and time are postulated to arise from this processing, with black holes just the grid processing matter at full capacity.

*Young's experiment*

Over two hundred years ago Young carried out an experiment that still baffles physics today - he shone light through two slits to get an interference pattern (Figure 3). As only waves diffract like this, light must be a wave, but then why do photons always hit at a point? When physicists sent photons through Young's slits *one at a time,* each still gave only one dot, but then the dots formed an interference pattern! The effect was independent of time, e.g. one photon shot through the slits each year still gives the pattern. As each photon can't know where the previous one hit, how can this occur? Can one photon go through two slits at once? In an objective world one could check it out, but in our world detectors in the slits just fire half the time, *never* both at once. In the quantum conspiracy of silence, a photon is a particle in one place if we look, but a wave in many places when we don't. This *wave-particle duality* is like a skier traversing a tree on both sides but still crossing the finish line intact (Figure 4). The problem is:

1. *If light is a wave*, why doesn't it smear over a detector screen as water waves do?

2. *If light is a particle,* how do one at a time photons create interference?

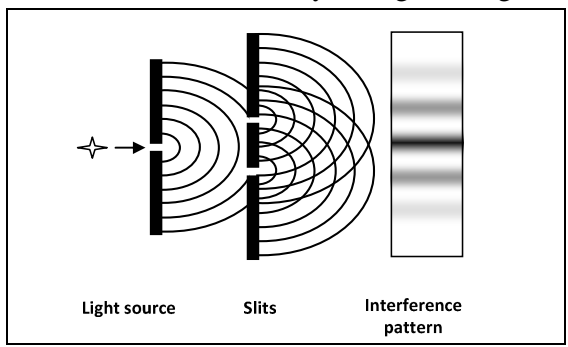

*Figure 3. Young's double slit experiment*

As Feynman said: "*all the mystery of quantum mechanics is contained in the double-slit experiment.*"

Quantum theory explains it like this: a photon has a *wave function* spreading in space whose power at any point is the *probability* it physically exists there. For each photon, this ghostly wave goes through both slits to *interfere with itself* as it exits, but if detected <u>anywhere</u>, collapses to a point, as if it always was a thing in one place. Quantum theory calculates the physical outcome using the wave equation to calculate values at every point, adding them to give a net amplitude, then squaring it to give the probability of physical existence at that point.

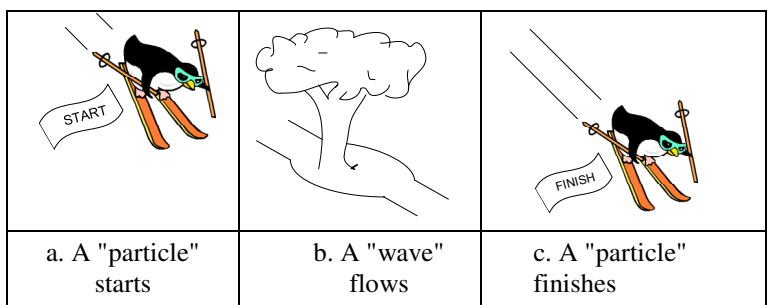

| a. A "particle" starts | b. A "wave" flows | c. A "particle" finishes |
|---|---|---|

*Figure 4. Wave-particle duality*

To see how strange this all is, note that the first photon to hit a screen in a *two slit* experiment is the first dot of what *will always be* an interference pattern. Yet if the first photon of a *one slit* study hits the same screen point, it is the first dot of what *will never be* an interference pattern. The outcome variation must be in the first events, but they were physically identical! Whether the slit a photon *didn't go through* exists or not decides if it is part of an interference pattern. How can a *counterfactual event,* that didn't happen, alter physical reality? This theory, of imaginary waves that conveniently collapse if viewed, is the most successful theory in the history of physics, but as John Wheeler succinctly said: *How come the quantum*? In other words:

1. *What are quantum waves?* What exactly spreads through space as a wave?

2. *What is quantum collapse?* Why does the wave collapse to a point if viewed?

In this model, a photon is a program distributed over the grid as packet instances, i.e. a spreading processing wave. It arrives at a detector screen as a "cloud" of program packets, each seeking processing from grid nodes already fully busy processing the screen's matter. As nodes have a finite capacity, they overload[13] and reboot, i.e. try to re-read <u>all</u> processing from scratch. The *one* grid node that reboots the *entire* photon program is the point where it hits the screen[14]. A reboot acquires <u>all</u> the instructions of the photon program for one cycle, so other instances disappear, i.e. quantum collapse is the inevitable disbanding of instances when a program source restarts [13].

The physical event of a photon hitting a screen then arises when many photon program packets cause many screen grid nodes to overload and reboot. The first packet to access the photon program source wins the quantum lottery of physical existence and gets to "be" the physical photon. This seems random[15] to us, as it involves program-node interactions we can't see, but one expects nodes running more entity program instructions to get more access, i.e. reboot successfully more often. To calculate a node's program access, first add the instances it handles and cancel positive and negative value calls at the node, as a program call efficiency. Then, as a sine wave's power is its amplitude squared, so the access demands of a processing sine wave will be its amplitude squared. The restart node probability, based on processing access, is then the square of the net amplitude. If processing waves can both add and collapse as quantum waves do, it is reasonable to equate them.

*The law of least action*

That a photon is a program of spreading instantiations impacts a problem that has puzzled people for centuries. Hero of Alexandria proposed that light always takes the shortest path between points.

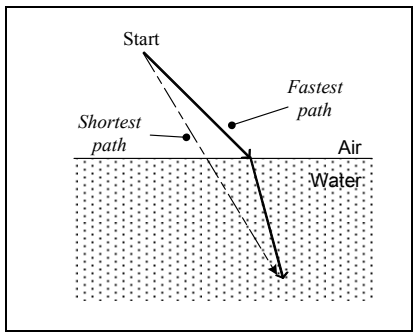

This is usually a straight line but in refraction it is not, e.g. the bent solid line that light takes in Figure 5 is not the shortest path. In 1662, Fermat amended it to be the path of least time. Imagine a photon as a life guard who runs faster than he or she swims, trying to save a drowning swimmer. It is quicker to run further down the beach then swim a shorter distance, as in the solid line, than take the dotted line shortest path, as refracting light does. So in 1752 Maupertuis proposed the *law of least action,* that:

 *"The quantity of action necessary to cause any change in Nature always is the smallest possible"*.

*Figure 5. Wave refraction*

Euler, Leibnitz, Lagrange, Hamilton and others developed this mathematically, sparking a furious debate on whether we live in "the best of all possible worlds". Despite Voltaire's ridicule, how a photon at each stage finds *in advance* the fastest route for all alternate paths, any media types, any complexity and inclusive of relativistic effects[16], remains a mystery today. To say it chooses a path *so that* the final action is less, gets causality backwards. As Feynman said:

 *"Does it 'smell' the neighboring paths to find out if they have more action*?" [14] p19-9

 Super-computers running a million-million cycles a second take months to simulate not just what a photon does in a million-millionth of a second, but in a million-millionth of that [12] (p113). How can a simple photon, with no known internal mechanisms, do all that processing work?

This model's answer is that multitudinous photon instances travel all possible paths, and those that *happen* to take the fastest route to a detector trigger physical reality. This idea supports Huygens principle, that *each wave front point is a new wavelet source expanding in all directions.* The resulting photon wave front then ensues as Huygens proposed, by reinforcement and interference, but what is distributed is processing not matter. This processing can restart at a point, i.e. become a "particle" on demand. The first detector grid node to overload and reboot successfully is where the entire processing wave re-spawns from. Quantum collapse is the garbage collection of other instances, like a clever magician removing the evidence of how a trick is done. When the photon processing wave reboots at an instance's node point, the latter's path *becomes* the physical path, i.e. *photons arrive in straight lines, they don't travel in them*.

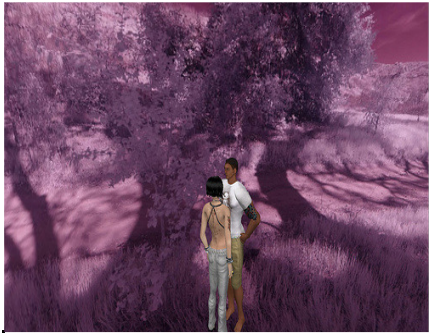

*Figure 6. A simulation*

How can a photon can know the fastest route to an unspecified destination unless it tries them all? In this model, quantum reality takes all possible paths and lets physical reality occur later, on a first come first observed principle. This *just-in-time* (JIT) computing might seem inefficient, but calculating and taking a path are the same thing in a virtual world. As the system *must* calculate every path anyway, to find the best one, this avoids double handling. If photon instances travel all available paths of the underlying grid architecture, Feynman's sum over histories describes exactly what happens [14] p26-7. A throng of instantiations try all the options and physical reality takes the best and drops the rest. In this *evolutionary physics*, quantum processing tries the variants and physical reality is the selection. The physical law of least action then implies the *quantum law of all action,* that everything that can happen in the physical world does in fact happen in the quantum world[17]. If it isn't the best of all possible worlds, it isn't for lack of trying.

*Relativity*

The grid proposed is not in space or time, *but what creates them*. It accommodates relativity, as both the fore-ground avatars and back-ground forest of Figure 6 are pixels. In computing, one can "move" a screen avatar through a screen forest by bit-shifting the avatar pixels with respect to the forest pixels, or by bit-shifting the forest pixels behind the avatar. If *there is no fixed node-pixel mapping*[18]*,* a centre-screen avatar "moves" if the forest scrolls behind, i.e. has a constant relativistic "frame-of-reference".

In this model, empty space is just the grid on "idle", like a blank screen that is still "on". Turning it off, to show the screen (grid) itself, would also destroy the images on it (our bodies). If quantum programs create the physical world can we hack into them? Quantum computing is doing just that already. When looking out a window one sees the glass by its flaws, by the frame around it, or by touching it. If the grid is a flawless transmitter all around us, it has no frame, and if it transmits matter as well as light, one can't "touch" it. Such a ground of existence can be deduced, as here, but not seen directly. If quantum mechanics describes the camera that takes the "photos" we call physical reality, it can no more appear in the pictures it takes than a finger can point to itself.

*Time as processing cycles*

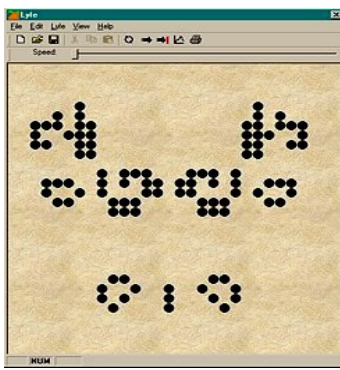

*Figure 7. Life,*
see http://abc.net.au/science/holo/lablife.htm

While objective time passes inevitably, by its own nature, virtual time is measured by processing cycles, e.g. in Conway's "Life" (Figure 7), pixels live and die by program events. A game of twenty minutes on an old computer might run in two seconds on a new one, but an avatar in both games sees the same virtual time pass, i.e. the same number of processing cycles. We measure our time in this way, as atomic clocks count atomic event cycles. In Einstein's twin paradox, a twin on a rocket accelerating to near the speed of light returns a year later to find his brother an old man of eighty because his time slowed down. In this model, the rocket's acceleration uses up grid processing, leaving less to process the life event cycles of matter, compared to his twin on earth.

By the same logic, for a photon what to us is time "stops", as the photon moves on before a node can "tick" a cycle. When people hear that our time can slow down they suspect a trick, but it is no trick. In accelerator experiments time really does slow down. It is not *perceptions* of time that change but *actual time,* as measured by instruments. Only in a virtual reality can this be.

### Space as the grid architecture

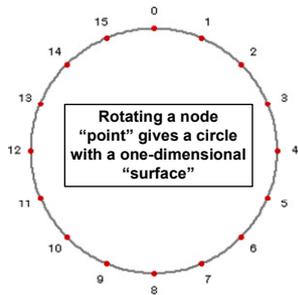

*Figure 8. A one dimension surface*

Continually dividing a virtual time gives a "tick" that cannot be paused, as continually dividing a virtual space gives a "pixel" that cannot be split. In our world, studying Planck length or time needs short wavelength light, which is high energy light, but putting too much energy in a small space gives a black hole that hides information from us. Probing the black hole with more energy just increases its size, to reveal no more. As closely inspecting a TV screen reveals only dots and refresh cycles, so closely inspecting our world reveals its resolution and refresh rate[19]. A digital universe of irreducible pixels and indivisible ticks resolves the *continuum problem* that has plagued physics since Zeno first outlined his paradoxes[20]. Loop quantum gravity, cellular automata and lattice simulations [13] map nodes to points in a *static* Euclidean space, but relativity requires nodes allocated to points *dynamically,* as the Internet allocates IP addresses - on demand. In this model, space is the grid architecture, i.e. how the nodes dynamically connect to each other.

### Creating dimensions

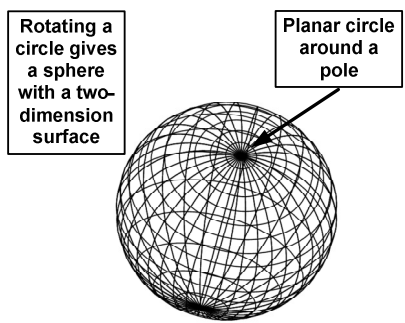

*Figure 9. A two-dimension surface*

Suppose a finite set of mini-CPU nodes arbitrarily link up into a circle (Figure 8). Each node has left and right transmit "directions", so there is one dimension. Repeating the notional rotation gives a two-dimensional sphere surface (Figure 9), with two dimensions of transfer movement. A *pole* of this sphere has a *planar circle* of nodes around it, where each node connection is a direction. Yet the rotation axis making that node a pole was arbitrary. A different axis makes another node a "pole" on the same sphere surface. As networks can easily alter links, let *each node configure itself as a pole* by setting its local links so. Now by rotational symmetry, *every* node has a planar circle of neighbours veridical to directions on an ideal sphere *with itself the rotation origin*. If each node can "paint" its own longitudes and latitudes, it can define its own space when activated, as relativity demands.

Figure 10 just repeats the above to give a 3D surface. A fourth dimension is hard to draw or imagine, but as rotating a circle gives a sphere so rotating a sphere gives a *hyper-sphere*[21]. It has a three-dimensional, simply connected, unbounded surface - just like our space[22]. Its *granularity* is the number of discrete steps in the N-rotations creating it, where a triangle is a 3-circle, a square a 4-circle, etc. Yet even with N large, discrete circles cannot tessellate a surface, i.e. this space has "holes" in it, so point entities like electrons could pass right through each other. In this model, that is not a problem, as quantum entities exist as instances across an area, so collide at a point if in the same vicinity. *That quantum entities exist inexactly lets an inexact, non-Euclidean space simulate ours.*

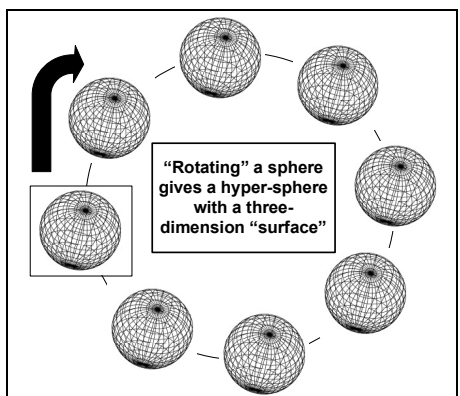

*Figure 10. A three dimension surface - cannot be drawn, only imagined.*

A node on a three dimensional hyper-surface has a sphere of neighbours, but in this model, *all grid transfers use a plane of our space,* i.e. occur in planar channels of a single rotation. This not unthinkable as in quantum Hall models two-dimension *anyon* excitations give quantum events [15]. If the number of points in a planar circle is the number of directions in that plane, there will be a minimum *Planck angle* for single node quantum events[23] [2].

## Implications

### Light as transverse processing

Does an extra dimension plus four dimensions of space-time give five dimensions in all? Not if time and existence are one and the same. This model has *four degrees of freedom*, three for space and one for *existence in time*. It matches the Hartle-Hawkin no-boundary theory, where one of four initially equivalent dimensions "somehow" became time rather than space [16]. Here, that somehow was as the dimension into which quantum wave entities oscillate to exist. Indeed, without an extra *space-like* dimension to curve into, how can a 3D space "curve", as relativity says it does? In this view, light waves move *on* our space as water waves move *on* a pond surface, i.e. they cannot leave it. We register them as one wave affects another, i.e. by existing in the same way. As 3D Flatlanders in a 4D reality, complex number theory needs an imaginary fourth dimension to explain light[24].

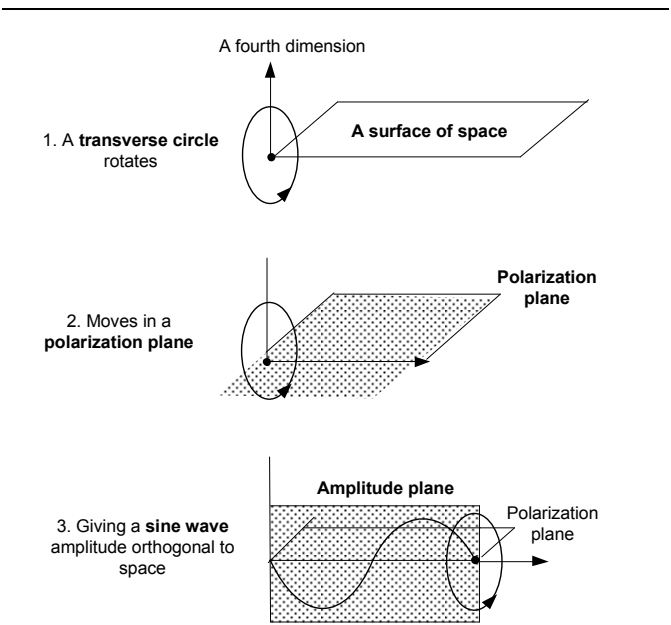

A fourth dimension

A surface of space

1. A transverse circle rotates

Polarization plane

2. Moves in a **polarization plane**

Amplitude plane

Polarization plane

3. Giving a **sine wave** amplitude orthogonal to space

*Figure 11. Photon sine wave projection*

In contrast, a photon as a three dimensional "thing in itself" poorly explains its wave behaviour. With no physical "ether", physicists just *declare* that light vibrates nothing[25], but vibrating nothing to give something makes no sense. Light as a "… *self-renewing field disturbance*." [12 p212] begs the question of what renews the fields that renew? That electric forces give magnetic forces that give electric forces, etc., is like Peter paying Paul's bill, and Paul paying Peter's bill. Were this possible, one could borrow money from a bank to pay back a debt owed[26]. In no physical pond do rippling waves suddenly become "things" when observed. Yet the "big lie", that light is a *wavicle,* is now physics doctrine, as Gell-Mann noted in his 1976 Noble Prize speech:

"*Niels Bohr brainwashed a whole generation of physicists into believing that the problem (of the interpretation of quantum mechanics) had been solved fifty years ago.*"

Do light waves oscillate in space as sound waves do in air? An objective realist might say how else could it be? Empty space transmits no sound but still transmits light, or we couldn't see the stars at night. Yet light is electro-magnetic, with positive-negative values. If space has no directionality, as "up" from one view is "down" from another[27], how can oscillations in space give positive-negative values? Or if something vibrates, why doesn't it fade over time, by the second law of thermodynamics? Light as a frictionless *wave of nothing,* vibrating in space, is quite implausible.

In contrast, information values can oscillate forever without friction. On a 3D hyper-surface, the plus-minus values of electro-magnetism are just in-out displacements, like the bumps and dimples on a ball, with the constant speed of light in a vacuum the surface refresh rate. Light then slows down in glass because the grid must process the matter as well as transfer the light. We *say* the medium of light is space or glass, but here the medium of light is *always* the grid[28].

The grid also keeps photons in sequence, like the baggage cars of a train driven by the same engine. If the engine slows down under load, as when near a massive object, photons go slower *but still keep the same order,* so no photon can overtake another. Were it not so, we could see an object leave then arrive! Our causality depends critically on photons keeping in sequence, which the grid engine rigorously maintains.

*The Planck program*

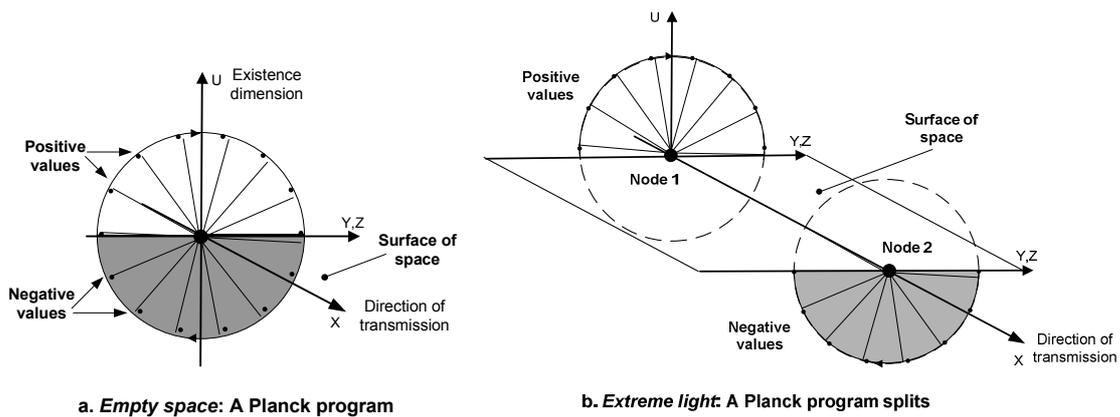

*Figure 12. Space and the first light*

In computing, a central processing unit (CPU) runs a program that tells screen nodes to set image pixel values. Here the grid is both "screen" and CPU, in our terms, as it both receives input and gives output. The basic grid "command" is one discrete rotation transverse to 3D space - a *transverse circle*. The instructions to turn a full transverse circle, a *Planck program*, projects the sine wave amplitude of light we see as it moves (Figure 11)[29].

Imagine a carnival wheel of black and white values spun by a pulsing machine[30] (Figure 12a). The pattern spun is the program, the machine turning it a grid node, and the result a quantum pixel state[31]. If a machine turns a full black and white pattern, the equal positive-negative values give zero, as equal up-down displacements on a surface cancel. One node running one Planck program per cycle is null processing, i.e. empty space.

If the same pattern divides over two machines (Figure 12b), each shows first white then black values, i.e. the effect is no longer null. Something now "exists", if only for one cycle. The *wavelength* of this extreme light is two grid nodes and its *frequency* is half the one-node rate of empty space.

The rest of the electro-magnetic spectrum (Figure 13) arises as the same Planck program increases its wavelength to divide among more grid nodes. If no instruction is allocated twice[32] more nodes share a program by running it at a slower frequency[33]. *Distributing processing runs it slower not less at each point.* In this model, light is the null processing of space distributed over many grid nodes instead of one. A photon then has zero *rest* mass because it is in fact "space on the move". If it ever "rested", for its wave train to catch up, it would become empty space again.

If *energy is the node processing rate*, a photon's energy comes in discrete packets because its wavelength is discrete, i.e. a higher frequency is one less node running the same program. So the highest photon frequency, with a two Planck length wavelength, must *double* its energy to reach the higher energy of empty space[34].

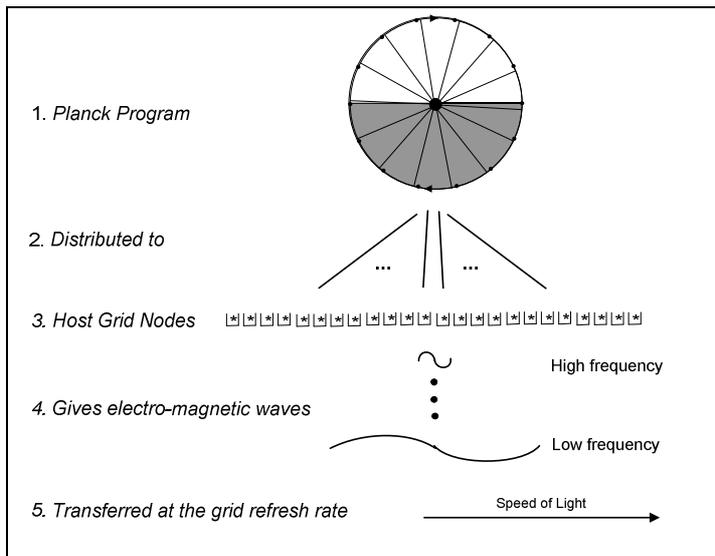

1. *Planck Program*

2. *Distributed to*

3. *Host Grid Nodes*

4. *Gives electro-magnetic waves*     High frequency     Low frequency

5. *Transferred at the grid refresh rate*     Speed of Light

*Figure 13. Electro-magnetism - a distributed Plank Program*

### The size of space

Plank's constant, the unit of photon energy, also defines the size of space*:* if it were smaller atoms would be smaller and if it were larger quantum effects more evident. Why does one constant define both energy and space? As proposed, Planck's constant represents the processing rate of one grid rotation per cycle, so it depends on the number of values in a *transverse circle*. As also proposed, a *planar circle* of connections defines the directions of space for a plane. The number of nodes in that circle defines its circumference, and hence the radius between nodes [2], i.e. the basic unit of distance. *If the grid is rotationally symmetric*, transverse and planar circles will have the same number of value steps. Planck's constant then links the quantum of energy and the size of space because it is the *granularity of the grid* that generates both

### The big bubble

An objective universe that "just is" may transform its parts but not its total *steady state*. So last century, big bang theory battled it out with respected physicists, for whom a steady state physical universe just "popping up" out of nowhere was highly unlikely. Yet that all the galaxies are expanding out from us at a known rate implies a source event over 14 billion years ago. Finding cosmic background radiation left over from the big bang has confirmed it for most physicists today.

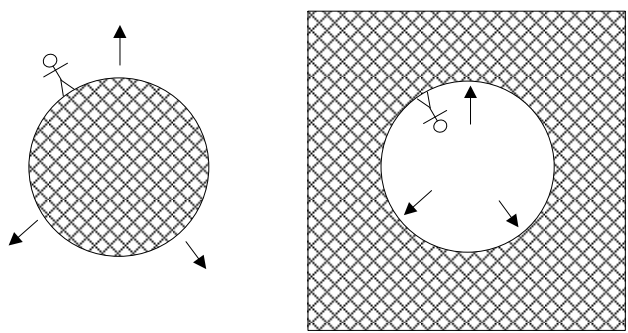

*Figure 14  a. A bang*          *b. A bubble*

The failure of steady state theory removed a cornerstone of support for the idea of an objective physical universe that exists in and of itself. Physics today still "defines away" questions like "*What existed before the big bang?*", but any universe that *began* is dependent, so what it depends on is a valid question. If time and space suddenly appeared, for no apparent reason, could they equally suddenly disappear today? If nothing in the universe comes from nothing, how can an entire universe do so? That an entire objective physical universe *arose*

*from nothing* is not just incredible, it is inconceivable.

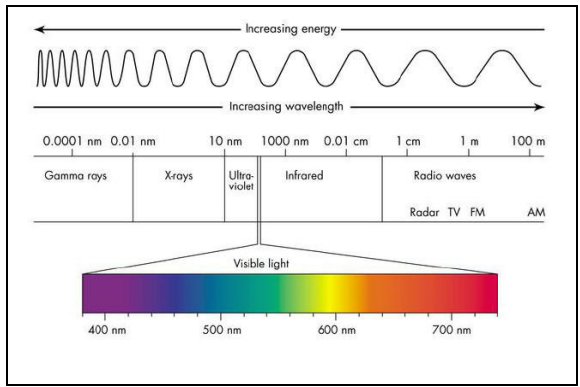

*Figure 15. The electro-magnetic spectrum*
from http://www.antonine-education.co.uk/

In contrast, this model *needs* a big bang. Every virtual reality begins with an information influx that starts *its* space and time. Any computer booting up starts a "big bang", before which there was "nothing" (in its memory). Here, the big bang was just when the universe booted up.

The term "Big Bang" puts us outside an expanding sphere (Figure 14a), but we could be inside a hyper-bubble expanding into the bulk around it (Figure 14b), whose inner surface is our space. A *hyper-bubble surface* is finite but unbounded and expands with no centre, just like our space. The expansion isn't evident as it doesn't alter existing matter - the bulk just fills the quantum "gaps" that arise everywhere, i.e. space itself expands. An explosion on such a surface will first go "out" then wrap around to end up everywhere. So cosmic back-ground radiation is still all around us, not at a some expanding edge, because it has circled our universe many times.

### The original event

If our universe collapsed in a "big crunch", it would soon form an enormous black hole. So if it was created all at once, at a point "singularity", why didn't the same immediately occur? In this model, there never was a big bang singularity. The initial event was *one* grid node "ripping apart" to give both the first photon and a grid "hole" across whose inner surface it moved as a processing wave[35] [3] p17. There was no black hole because it all began as *one* photon in *one* volume of space. An unimaginable cataclysmic chain reaction followed, as the grid tore itself apart to create a universe of quantum waves on the surface of space. The rip spread like a tear in a taut fabric[36] but the expanding surface it produced weakened the vibrations enough to stop the ripping, though space still expands today. So *inflation*, when space expanded faster than the speed of light [22], was *the grid itself breaking apart to create our universe.* The free information of the universe has been constant since then, as the grid today no longer tears. By the expansion of space, the original vibrations descended into lower and lower frequencies, from x-rays to radio waves to create electro-magnetism (Figure 15). That this processing also created matter is now proposed.

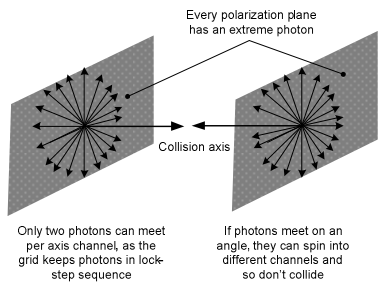

*Figure 16. Extreme light meets*

### The matter glitch

If the total processing requested photon packets meeting in a planar channel is less than its bandwidth, of one Planck program of instructions, it will not overload. So photons normally pass through each other and do not collide. The exception is *extreme photons,* with a wavelength of only two grid nodes (Figure 12b), that set the Planck program's "up" values in one node and its "down" values in another. Two such photons meeting head-on in a grid channel will set all its values, i.e. overload it. A photon that *spins* on its movement axis can switch to a different channel [3], but what if every axis channel has an extreme photon? If *extreme light*[37] meets, every channel overloads with no alternatives (Figure 16). Figure 17 shows what happens when two extreme photons meet head-on in one channel - it overloads and reboots, i.e. re-starts all its processing. The reboot first passes all processing back out to its channel neighbours, but then the photon tails also overload, giving another reboot. This recurs endlessly, as repeated measurements can collapse a quantum wave system back to a constant state [23], [24] p23. The grid fabric that once hosted only dynamic undulations now has a permanent processing "bump" - an electron. The constant processing remainder, of negative Planck program instructions, is proposed to be its charge, i.e. *charge is the system keeping its processing books.* An electron is when a grid node gets in an infinite loop, i.e. "hangs".

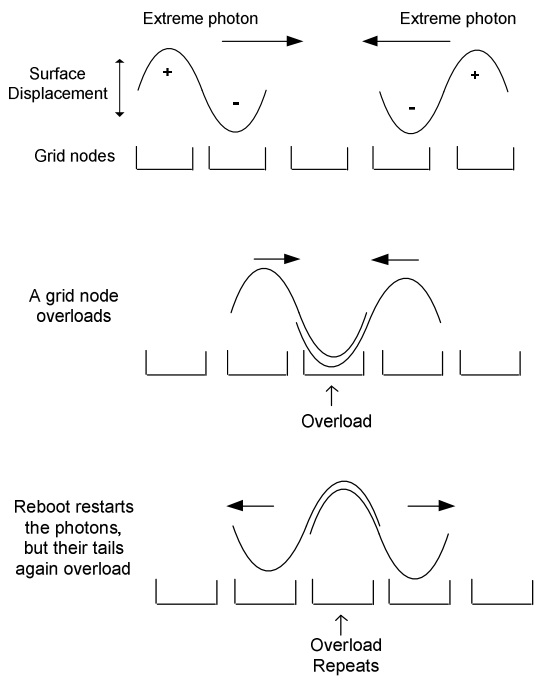

Without electrons there are no atoms, life, or us, but why does the universe have neutrinos, little "nothing" particles with almost no mass or charge? In this model, neutrinos occur when the photons of Figure 17 meet *out of phase.* Now the processing cancels to a recurring smudge on the surface of space (Figure 18), i.e. a neutrino[38]. If electrons and neutrinos arise in the same way, a universe of electrons *must* have neutrinos. It is shown elsewhere that extreme photons colliding on *three axes* can give up and down quarks, and their fractional charges, again by phase [4]. In this model, all matter is a processing glitch.

*Anti-matter*

As the equations of physics are all reversible[39], physicists have often wondered, why isn't time so? In this model, time is not a sequence of self-existing physical states[40] but the sequence of dynamic choices that create them[41], i.e. time is processing. A processing sequence can run in reverse only if all the choices that constitute it are reversible. Clearly quantum collapse, as a processing restart, is irreversible, because a reboot by definition loses all previous data. If all physical events are processing reboots, they cannot be reversed.

*Figure 17. A grid channel "hangs" (electron)*

However *between interactions,* the "add-one" processing of virtual existence can be reversed[42], allowing an "opposite" existence, i.e. anti-matter. If matter sets values clockwise *with respect to its movement on the surface of space*, anti-matter sets the same values anti-clockwise. Processing electrons and neutrinos in reverse gives their anti-matter equivalents (Table 1). As can be seen, electrons are symmetric in space but neutrinos are not, which explains the puzzling "handedness" of neutrinos, that turning an electron in space doesn't alter it but turning a neutrino does. If the difference between matter and anti-matter is processing cycle direction, the node split that began the universe was the "symmetry breaking" that gave a matter rather than anti-matter universe[43].

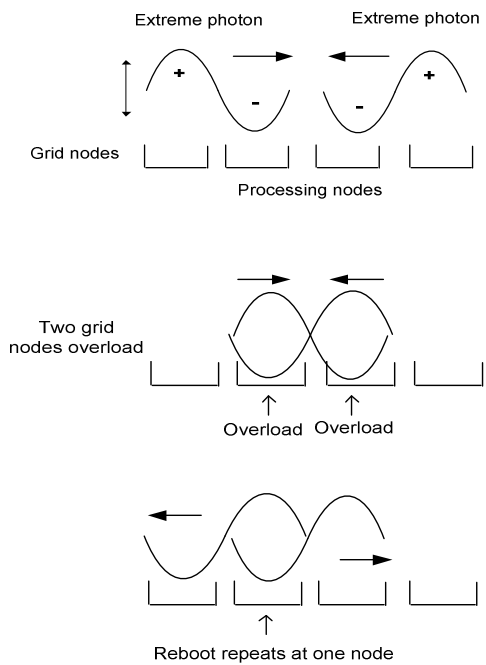

*Figure 18. A grid channel "hangs" (neutrino)*

Why then in Feynman diagrams do anti-matter particles enter events going backwards in time? In this model, anti-matter can no more reverse a causal interaction than matter can, i.e. interactions can't be reversed.[44] But if anti-matter *exists* by processing the same values as matter does in reverse, by definition again, it runs *our virtual time* backwards. Of course in its virtual time, it is our time that is running backwards.

*Matter and energy*

Einstein's mass-energy equation gives energy and mass equal status. It says they are the same thing, not that both are mass. The equation by which nuclear matter becomes energy also lets energy become matter, i.e. *Einstein's equation works both ways*. As metric and imperial both measure the same thing, mass and energy measure the same thing, so why does the standard model of physics treat everything as matter-like particles?

A continuous world of objects that inherently exist is a concept flawed at its foundation. If every object has a continuous substantiality, it must be divisible into parts. Each part then needs further parts and so on, but if every object is composed of sub-objects, how can it ever end? The standard model purports to offer "fundamental particles", not constituted of anything else, but then lets them transform into each other, e.g. electrons and positrons annihilate into photons. How are particles that transform into each other "fundamental"? That a top quark, whose mass equivalent is a gold nucleus of 79 protons and 118 neutrons, is a fundamental particle with no internal structure, makes no sense. Why then does it decay into lesser quarks? Thinking physical objects are always constituted of smaller physical objects is like thinking the earth sits on the back of a giant turtle. As that turtle would need another turtle to stand upon, ad infinitum, so every object needs sub-objects to comprise it. The world can no more be "objects all the way down" than "turtles all the way down"[45]. The existential buck must stop somewhere, and in this model, *processing is it*.

| | **Electron** | **Neutrino** | **Positron** | **Anti-neutrino** |
|---|---|---|---|---|
| ***Structure*** |  |  |  |  |
| ***Charge*** | -1 | 0 | +1 | 0 |
| ***Key*** |  | | | |

*Table 1.  Leptons as extreme light collisions, showing one channel only*

If mass and energy are forms of processing, and energy is the node processing rate, what is mass? Mass, unlike energy, doesn't change over time. In a program loop, as matter is here proposed to be, what doesn't change is the program that run every cycle. If mass, in processing terms, is *the number of program instructions read per node per cycle*. Einstein's mass-energy equation then follows directly[46].

*Light colliding*

In current physics, light and matter[47] are different existence types, so:

"*Two photons cannot ever collide. In fact light is quantized only when interacting with matter.*"[3]

In contrast, this model predicts that *extreme light colliding in empty space can give stable matter*. That matter is *light bottled up* contradicts the standard particle model of physics. We know that photons can give temporary electron-positron pairs by the Breit-Wheeler reaction, that accelerator collisions can create new matter[48], and that high-frequency light near a nucleus can create matter, but to predict that pure light can, in the extreme case, collide in a vacuum to give permanent matter is a shock. If matter arises from light, it is not fundamental. Physics framed around the nineteenth century particle concept still dominates, even though space creates matter, matter plus matter creates matter and light plus matter creates matter, but if light alone in a vacuum can create matter, it must fall.

---

[3] Accessed August, 2010 from http://en.wikipedia.org/wiki/Two-photon_physics

## Conclusions

In conclusion, the VR conjecture is neither illogical, unscientific or untestable, nor does it deny a real world "out there". To make the physical world the virtual output of a quantum reality at first seems outrageous. Yet it resolves the main problem of modern physics: that its primary theories, relativity and quantum theory, contradict not only common sense but also each other:

1. *Quantum theory* assumes fixed space and time backgrounds[49], which relativity specifically denies. For quantum theory to satisfy relativity, it must be *background independent,* i.e. not assume, as it currently does, that quantum states in a fixed space evolve in a fixed time [27].

2. *Relativity* assumes objects exist locally, which quantum theory specifically denies. For relativity to satisfy quantum theory, it must be *foreground independent,* i.e. not assume, as it currently does, that local objects travel fixed trajectories.

These theories clash because each denies an objective reality assumption the other has ignored[50]. In the VR conjecture, "space" is the grid channel architecture, "time" is the processing cycles of a grid node and "objects" are distributed processing waves, i.e. *neither foreground nor background are absolute*. Everything is dynamic processing, whether time, space or objects.

By the logic of quantum theory, between our "real" observations is a quantum unreality of which the Copenhagen doctrine says *we must not speak*. Yet as quantum entities are in-between interactions more than in them, the world exists mostly in uncollapsed quantum states. If so, by what logic are its brief moments of collapse "reality"? *Surely reality is what exists most of the time?*

Or if quantum waves predict and cause physical reality, isn't making a cause "unreal" and its effect "real" a backwards logic, like saying the sun circles the earth? If quantum states create physical states, how are they unreal[51]? *Surely reality is that which causes, not that which is caused?*

The denial of quantum reality by current physics is doctrinal not logical[52]. It contradicts the evidence that quantum states and quantum collapse are non-physical by nature. When matter was first attributed to unseen atoms, scientists like Mach denied it. Now atoms have electrons, protons, neutrons and even quarks that are never seen. Yet when quantum theory finds answer to the universe and everything is a numerical probability[53], we say "*Stop! Enough!"* That the physical world is the output of a quantum calculation seems a step too far. Yet if so, it would not the first time science has told us what we don't want to know.

It is to this place, that others avoid, the VR conjecture takes us, not to shock or amuse, but to advance. We assume ourselves in the rational sunlight of physical reality, standing before a dark cave of quantum paradox, but in this model, as in Plato's cave analogy, it is the other way around. We sit in the dark cave of physicalism with our backs to the quantum sunlight, observing a shadow world projected on the wall of space. That quantum processing gives physical reality as a calculator gives a total seems inconceivable, but it explains why, in quantum theory, physical reality only exists if we look. One must query the quantum database to get a "view". What we see then is, as Kant said, not *what exists in itself,* but what the quantum firewall presents.

In this model, a photon is a program, its movement a grid transfer, and a physical observation is a grid reboot. The test proposed is that the highest frequency of light will collide in a vacuum to create permanent matter. To test this, physics must stop smashing matter together and start colliding light. If extreme light rays collide to give an electron, as this model predicts, then physics must rewrite the "story" behind its equations, as a story of processing not particles. If processing creates physics, then physical realism will be dethroned, but mathematical realism will be restored, i.e. quantum equations now refer to something real. If science tells us that the physical world is digital at its core, we may need to rethink our position in the scheme of things.

## Annex A: Discussion

*Comment:* We start from opposite axioms: I claim that we cannot go outside of the physical universe for physics - all else is equivalent to an appeal to God (which is legitimate but not physics). By contradicting this "prime axiom", that "There is nothing outside the physical universe", the floodgates are open to let anything convenient through, no matter how unlikely or even absurd. You deny objective reality. What is surprising is what a convincing argument you make, but I believe a number of your arguments are either wrong or misleading.

*Author:* It is just a conjecture, a question not an answer, albeit one that needs airing. This is not a *theory* of everything (TOE) but a *query* of everything (QOE). When I started this, I also thought it would soon fall apart, so am equally surprised it hangs together. Re that "*the floodgates are open to let anything convenient through*" - it isn't so, as long as we stay scientific, i.e. use consistent logic, collect data, test predictions, etc. This is a theory about <u>this</u> world, not any imaginary metaphysical world made up. Its basic idea is that the physical world is a processing output. If so, we can reverse engineer it, i.e. apply the "grounded theory" method of sociology to physics. It doesn't advocate (or deny) new age or religious ideas, i.e. you can still believe what you want to believe [29]. If some use these ideas to justify their pet theories - that's life isn't it?

*Comment:* You begin, as many do, by denying local realism. That is the current trend, probably because it's 'sexy'. But one of the world's foremost experts, Anton Zeilinger, has written a book, Dance of the Photons, where in Appendix A he proceeds to derive Bell's inequality and to claim that actual measurement results imply that the properties "do not exist until measured". And here is the catch, the entire logic is based upon the assumption that the properties do not change en route to being measured! If this assumption is wrong, then the logic of Bell's inequality is wrong, and the drastic step of denying local realism is simply not justified.

*Author:* If the properties of a photon change *en route*, without physical interaction and before it is observed, isn't the objective reality hypothesis conceded? That a physical photon "thing" can change its properties for no physical reason, is indeed a floodgate. So I support Zeilinger. By D'Espagnat, Bell's logic assumes *physical realism, Einstein locality* and *logical induction* [30]. The experimental results mean that one of these three *must be wrong*. The VR conjecture resolves this by moving the word "physical" from the definition realism to that of locality:

1. *Realism* is that there is a (deleted the word physical here!) reality whose existence is independent of human observers

2. *Locality* is that no <u>physical</u> (added the word physical here) influence of any kind can propagate faster than the speed of light

This model drops universal locality but keeps physical locality, i.e. limits Einstein's logic to physical objects. It also drops physical realism but keeps realism, i.e. permits a non-physical quantum reality. The reason to give up "localism" is not flimsy but by the results of Bell's experiment [31]. The assumption targeted is specifically that two photons separated in space are necessarily two local "things". In this model, when photons entangle they merge into one program with a total spin of zero. It is non-local, as a program can make a black and white pixel pair anywhere on a screen [3] p 29.

*Comment:* You say "If the properties of a photon can change en route, without physical intervention... isn't the objective reality hypothesis conceded?" but I am not proposing "without physical intervention". In that case there is no violation of Bell's inequality. It is only violated when the photons are treated differently by polarisers or beam splitters, and I consider this "physical intervention". If the choice is to give up local realism or to believe a beam splitter has a physical effect on a photon, the choice is easy.

*Author:* I don't think Bell's logic is flawed. Take a simple case: a Caesium atom sends off two photons in opposite directions with unknown spin. Define "en route" as being from that source event until either photon is involved in any physical event interaction. So travelling through space, air or glass is "en route", but any physical or quantum interaction, like measuring its spin, is not en route. So if the apparatus has any physical effect, the photon is no longer en route. Yet it could affect a photon's "hidden properties" which become evident later when spin is measured, as Einstein thought.

As spin is a binary outcome with only two values (clockwise/anti-clockwise) a photon's probabilistic changing would be to entirely reverse spin as it moves. This is not like some cork bobbing on a quantum sea, but rather a trap that irreversibly snaps shut up or down if anything touches it, and does so randomly, i.e. regardless of prior physical events. In entanglement, two such traps set off in opposite directions and if one snaps shut "up" the other is always "down". If the "apparatus" causes these changes, why is this balance kept? Entangled photons can travel light years in empty space before a measurement, so how does the isotropic "apparatus" of space keep them always opposite? The two-slit experiment also implies non-locality, and the vacuum is again the "apparatus". In the VR conjecture, a program is always "non-local" to its screen, i.e. it can change any pixel anywhere on the screen equally easily. A program can keep two pixels in a black/white relation no matter how far apart they are on a screen.

*Comment:* If properties do exist they are expected to as you say: "ensure a constant spin zero" or "keep one black if the other is white", but then there is no necessary 'non-localism' since the existence of conserved properties means that if one is known, then the other is known. There's no need for 'spooky' communications between Bob and Alice's locations. Why is this not obvious? Because the Copenhagen 'superposition of states' inherently does away with realism in favour of mysticism, claiming that quantum objects are 'ghostly' until measured.

*Author:* It only does away with physical realism, not realism per se. When you say we still do not know what a quantum field is I add "physically". We know what it is mathematically. If one defines postulating anything "beyond the physical" as "mystical", then quantum theory and most of modern physics is mystical, including string theory. The VR conjecture just expresses this bluntly, without a "cover" of mathematics. Physics left the enclave of positivism long ago, when it embraced the idea of a field, because we don't "see" fields, just their effects. As Feynman said: "*A real field is a mathematical function we use for avoiding the idea of action at a distance.*" [29] Vol. II, p15-7 . What is action at a distance but non-locality? That the physical world is virtual lets it be non-local, as a program can immediately alter pixels anywhere on a screen (even if the "screen" is all space!) I can't add to the Bell experimental evidence on this, except to agree with their conclusions.

*Comment:* As to your point 8: "Superposition. Objective entities cannot spin in two directions at once as quantum entities do...". The physical fact is that a magnetic field can only be measured along one axis at once, and this has been distorted by probabilistic representation into spinning in two directions at once.

*Author:* The physical fact is that a magnetic field is only measured on one axis at once but the "quantum fact" is that it can go both ways at once. This is why quantum computing is so powerful compared to physical computing. Quantum theory is not that we don't know how a photon spins and so describe it by probabilities, it is that it divides its existence into *actually spin* both ways at once.

*Comment:* Would you care to elaborate a bit on the differences between cases 2b and 2d? They seem similar, although 2d has two observers (or perhaps that's a single reflexive observer).

*Author:* Sure. In *physicalism* (2a) the world is as it seems: solid, real and self-existent. Consciousness emerges from physicality at about the complexity of the human brain [10], so machines will soon become conscious and replace us. Yet computers are socially dumb [32], as their Von Neumann architecture denies the self-awareness an "I" needs [33]. Piling up video boards into a supercomputer, like piling up rocks, just makes a bigger heap of the same thing. So physicalism is dominant but struggles to explain human consciousness.

Enter *solipsism* (2b), that we dream an unreal physical world. In psychology, every idea is mediated by the brain, so it could create a dream world[4], with no actual external world out there. If so, how is it consistent between observers? Or if it is *all* in my mind, then "you" don't even exist, i.e. the theory doesn't generalize well! Or if "I" am "All", then why does the world contradict my expectations? If the mind creates the body as in a dream, why can't I dream the body I want? If no-one observes a forest no tree can fall in it, but what if one looks later to see that a tree fell before? Is a consistent past history made up? Did our minds also fabricate the dinosaurs of history? In weak solipsism, human consciousness dreams the world by triggering wave function collapse, as Wigner suggests, but if humanity is the "observation central" that collapses quantum waves, how did the universe manage before we arrived to observe it into existence? Solipsism, that our minds somehow create the world, in its various forms, struggles with the realism, that a real world exists apart from us "out there".

---

[4] e.g. Harunyahya's "The secret beyond matter" http://www.youtube.com/watch?v=X04jN_xcLis

*Dualism* (2c) just merges 2a and 2b, to get the benefits of both, but also the problems of both.

*Virtualism* (2d) is the true opposite of physicalism (2a) not dualism (2c). As physicalism postulates a consciousness emerging from a physical substrate, so virtualism postulates a physical world emerging from an observer substrate. While it at first seems bizarre, recall that before the big bang, the entire physical universe didn't exist. So why is physicality the prime "reality"? In this model, an original quantum existence "observed" itself, and in doing so generated physical reality, i.e. the universe is more a big thought than a big machine.

In 2d, unlike 2a, consciousness is not a recent acquisition but what began it all. There are no "gaps", or view histories, to recapitulate if a quantum reality has simulated itself to itself from the start. While in 2b conscious observers initiate quantum collapse, in 2d observer and observed are symmetric, i.e. a photon observed is equally an observer. So no tree falls in a forest unseen - the ground it hits "sees" it. The virtual reality arises as a non-physical quantum reality "knows" itself. So observations create the physical world, just as quantum theory says. As in a virtual reality game, if I cursor left a left view is created, but if I cursor right another view is shown. This cracks the quantum measurement problem 2a faces, but unlike 2b there is still a real world "out there" - it just isn't the world we see. What is real is the observer, not the observed. In sum:

a. *Physicalism,* that the world just is, struggles with modern physics and consciousness.

b. *Solipsism,* that I dream the world, struggles with realism and doesn't generalize well

c. *Dualism,* that both the above are true, is an illogical compromise that solves nothing.

d. *Virtualism,* that a quantum reality generates a virtual physical world, is bizarre but possible.

In the latter, *every physical thing arises from consciousness.* Conway's theorem [34] proves that *either* we and the world are mechanical, *or* both we and it have free choice [3], i.e. if we are "conscious" then so is every electron, or if an electron is not then neither are we. So the monisms of physicalism and virtualism are the only valid contenders for a consistent universal theory. The physicalism case is Everett's many worlds theory (MWT), that every quantum choice spawns an alternate universe [35]. This contains the quantum ghost in a clockwork multiverse, where no choice is ever really made. Yet why should a multiverse, like a doting parent with a video-camera, copy an entire universe for every quantum event of our world? Why are we, *again,* the centre of things?

In contrast, in virtualism the quantum ghost is real and the physical world its virtual shadow. Science can test the hypothesis by comparing alternatives: the world is an objective reality vs. it is a virtual reality. So physics now has two consistent but bizarre explanations of its facts, MWT and virtualism!

*Comment:* Why couldn't an immense multiverse copy everything our universe might do? After all, if "our" universe and "the" multiverse are disjoint, there isn't anything to prevent replication and a subsequent new initial condition for "the" history of a subset of the multiverse, independent of "our" history.

*Author:* There is no reason it couldn't be so. MWT is THE physicalism explanation of quantum mechanics. While initially ignored, physicists today prefer MWT three to one over the Copenhagen view [36] p6. Yet if science picks the simplest alternative, the question is not if MWT is possible but if it is likely. MWT offends Occam's razor by assuming more than it explains, as the: *"… universe of universes would be piling up at rates that transcend all concepts of infinitude*." [37] p107. Deutsch's attempt to rescue MWT by letting a finite number of universes "repartition" after each choice, just recovers the original problem - what decides which universes are dropped? [3] p31. MWT was made up to support a bias and adds no value apart from that, to theory or practice. The VR conjecture is simpler, as it only duplicates quantum entities by program instantiation, not entire universes!

*Comment:* Your causality 'grid engine' keeping photons in order collapses instantly with Einstein lensing as the discrete field model of reality (DFM) predicts long lensing delays, e.g. we have recently found over 3 YEARS delay, i.e. photon 2 was emitted later but arrived here on earth over 3 years before photon 1.

*Author:* In gravitational lensing, photons from the same cosmic event that take different cosmic paths arrive on earth at different times. That doesn't collapse the idea that photons keep in lock-step order for a given path. Different paths will take different times if there are more or less transfers. Even paths of the same length will take different times if they involve different grid loads, e.g. if a path goes by a massive galaxy it will slow down the photon stream. This is also explained by general relativity in terms of space "curving" rather than processing load.

*Comment:* I propose the universe is real and simpler than we thought because our conceptual pattern matching skills are as yet undeveloped. Non-locality is just a misunderstanding

*Author:* The argument for non-locality doesn't depend on our pattern matching skills. It is ok to speculate, as physics does all the time, if it is worked out in detail, generates new knowledge and is logically consistent. We should be consistent, e.g. Einstein said that quantum theory denies empty space, as if an electron quantum wave spreads over a galaxy, how is space empty? This model has no empty space, because the grid fills everything and is *always* processing. It is the *medium* of physical existence, a non-physical "ether" whose output is physicality. But to be consistent, one can't have many grids. So this model must explain all the fields and conservations of physics as aspects of grid processing, e.g. electrical, magnetic, gravity, strong and weak forces. In contrast, any alternative theory must explain how the vacuum of space has effects, like the Casimir force.

*Comment:* I agree with many things you say, arguments you use, questions you ask. But by drawing a conjecture, you are not required to really look for an answer, just to speculate about it. I like your illustrations and your long bibliography. If you were to find the ultimate understanding of the universe, it would be so radical that, most likely, you would have no bibliography and would have not only to explain it but also train the reader into thinking differently.

*Author:* That a scientist must do more than philosophically speculate is why string theory is criticised for predicting nothing. This theory accommodates past unexpected results but must also develop to predict new ones. It is difficult to contradict past conditioning, that the physical world is a self-existent reality, but it was once equally self-evident that the sun went around the earth and that view changed. Ideas evolve as people evolve, which training can encourage but not externally impose.

*Comment:* Maybe this is nitpicking over terms but when you characterize VR as a simulation do you mean emulation? A simulation to me is the virtual reproduction of a model that actually exists physically while an emulation cannot at all be distinguished from the original. If all is virtual rather than physical, there is no way in principle to know if the emulation we experience in our " ... *dark cave of physicalism, with our backs to the quantum sunlight, watching a shadow world projected* ..." is original, or the emulation of an emulation, of an emulation, etc.

*Author:* An emulation mimics an original, as in emulating Windows on a Mac. Simulating can mean the same, as in traffic model simulations, but doesn't have to, e.g. in "The Sims" or Civilization a *world in itself* is created with its own space and time, that does not "realistically" copy its containing world. This is our case, as physical properties are not quantum properties, and the quantum world described by our mathematics is not at all like the physical world we see. Yet they connect, e.g. a VR, being contained, must have fewer dimensions than its parent, e.g. we view 2D screen surfaces in our 3D space. So if our 3D space is a "surface" of a 4D grid, perhaps the latter is a surface in another 5D reality, who knows? Science can only address this world, as it is all we can test.

*Comment:* I find more to agree with in your hypothesis than to disagree. MWT does not add value to a physical theory beyond the assumption that the world is objective. I don't know if the attribution can be verified, but have heard Hawking quoted as saying the MW interpretation is "trivially true." I agree, whether he said it or not. Point is, I'd need more convincing that my assumption of an objective reality is superfluous. As it stands, I don't have to accept any reality at all in order to do science. The assumption of objective language is sufficient, which is why I prefer "metaphysical realism" (Popper) to your term "physicalism."

*Author:* That science can theorize beyond the physical world but must validate from within it may be what Popper means by meta-physical realism. If so, it is a *method* of modern science while physicalism is a *conclusion* reached, i.e. there is no contradiction as they are different things. This theory is about what quantum theory means, not how it was derived. In an objectively real world, extra dimensions must be within it so string theory makes them "curled up", too tiny to see. Why an objective world would create unseen and unused dimensions is unclear. In contrast, in this conjecture we are the 3D version of Abbot's 2D Flatlanders [14]:

1. *Mr A. Square* imagines a "sphere" by linking his world's circles in a new dimension.

2. *Complex number theory* imagines a fourth dimension, apart from our "real" space, to describe electro-magnetic sine waves.

The physical world is not unreal, as in nihilism, but just *locally real*, i.e. with no inherent reality. This is not a religious or philosophical view, but a logical model implication. As simulations only exist by the programs creating them, so the physical universe requires quantum processing. If that processing were at any time to stop, the entire physical universe would, by this model, <u>instantly</u> disappear, as if it had never been. Clearly, it is a different world view from that of a self-sufficient physical universe, permanently existing in and of itself.

*Comment:* I do not comment as a physicist or mathematician, so you may ignore my comments freely. My pile was made higher and deeper long ago in engineering, but my life long study is metaphysics. I feel you did an admirable job of describing the world that we know as "around us" as illusion and the logical reasons to view it as so. In present day terms arriving at the same result as Siddhartha Gautama and other metaphysicians did so many centuries ago: that the world is illusion, or in today's terms, a virtual reality. I only find you to fail in not taking that step too far, for today's academic world, in not stating the source to be Mind/Consciousness.

*Author:* The term consciousness is currently ill-defined, so people take it to mean as they want, e.g. as based on sense *qualia*. So that I see red, or feel happy, is called "consciousness" in western science. In contrast, in Buddhism consciousness is nothing to do with the discriminating brain of the senses. Hui-Neng, the Sixth Patriarch of Chan and Zen Buddhism, spoke of *essence of mind* not ordinary mind. The traditions of Buddhism and western science are not the same, i.e. their "consciousness" doesn't refer to the same thing, so it is difficult to talk about it successfully. In this model, consciousness is the ability to observe, to be an information destination (*sink*) or origin (*source*).

*Comment:* It is a pleasure to read your paper but I think it is premature to pursue the VR conjecture at this point in time. Most, if not all, physicists would agree that "*The quantum world is in every way physically impossible, so physicality cannot be the nature of its reality*" but what if Einstein was correct? It is premature to assert that "*If science finds that it cannot be objectively real, it must explore if it is virtual.*"

*Author:* Well maybe, but physicalism, has had a 100 years to explain quantum theory. All it has come up with is the Copenhagen view that meaning doesn't matter, and the fantastical many worlds theory. There has also been plenty of time to find Einstein's hidden variables. So how long should they be given, another 100 years? Physicists might be happy with formulae, but people want meaning. Why does it hurt to explore another option? This essay is speculative, but supports its speculations with facts. As a method to test the theory is outlined [3], it deserves a chance, as all theories begin as speculations, e.g. atomism did. This is a theory about the world we see, that it is an information processing output, not a permanent, self-existing objective world. The properties of our world are then considered in turn, that it began, that it has a maximum transfer rate, that its "empty" space is not empty, that its time and space are malleable, that particles teleport, that it is discrete at the Planck level, etc. In each case, our world behaves as a virtual reality would, not as a world of solid "things" would. Why then is it "way-out" to conjecture that it a processing output?

*Comment:* You see "*a world created entirely by information processing as an output*" and hope to "*reverse engineer*" it. In my considerable hardware and software design experience the probability of success in such a venture is vanishingly small. I believe that I can design a counter example [I'm not offering to do so] that would use entirely different architectures [including analogue, digital, and mechanical parts] that would provide identical output, thereby preventing even the possibility of such reverse engineering. For the moment I'll just state my professional opinion that reverse engineering is not feasible and almost certainly not possible. But that's just my opinion.

*Author:* Most computing experts would I am sure agree with you, which is probably why it has not been attempted in detail before. Reverse engineering is to infer processing based on the input/output we see, e.g. in Jackson Structured Programming input and output *define* the processing between them. While this *may* be impossible for the world, in my experience one doesn't know what *is* impossible until one tries. Had I known ten years ago how hard it was, I might not have begun, but I just did. The aim is to reverse engineer the physical world [1], to specify how processing can output time, space [2], light [3] and matter [4]. All I can say is that so far it seems to be working. Reverse engineering is not a model postulate but a way to generate a testable model. Unlike creationism, it deduces a *program* from an output not an all powerful cause, i.e. it specifies the processing that satisfies quantum equations. In contrast, intelligent design involves no program specification so it isn't reverse engineering. It's a completely different situation.

*Comment:* You say "*The VR conjecture is not a theory about God or one that requires a God to exist*" but I find a theory of something outside our local reality (in an 'extra' dimension not perceivable by us) that accounts in an unknown way for what we see, is indistinguishable from speculation about God. What is the difference between:

1. God creates everything that we see

2. Information processing creates everything that we see

If there is no hope (as I contend) of discovering the architecture (of either) by reverse engineering, it is no different from conjecturing: "God creates it." I want to make it absolutely clear that I am not saying that you are postulating God. I am saying that, if reverse engineering is not possible, then you might as well be postulating God, because there would be no physical or logical testable difference between your theory and one that postulates God.

*Author:* If one defines a "God theory" as one that references beyond the physical world of our senses, then by a circular argument the VR conjecture can be categorised so. Yet postulating beyond what we see does not postulate God, e.g. Many Worlds Theory (MWT) postulates beyond what we see but does not postulate a God, nor does Bostrom's simulation hypothesis [38], or string theory with its many dimensions "not perceivable by us". To tolerate string theory's many extra dimensions but not VR theory's one extra dimension is a double standard. Every idea that denies positivism is not necessarily a "God Theory". Nor is it *for all practical purposes* (FAPP) so, unless MWT is the same, which few would see it as. The VR conjecture is about this world we live in, not about God. It is Tegmark's "physics from scratch" with everything derived from processing. If this works, it implies something beyond the physical universe, but doesn't presume to define it. Who knows what it is? It could be a Big God or a Big Machine - take your pick based on your world view bias. This is NMP (not my problem). I leave it to the theologians and philosophers.

*Comment:* You say that Many Worlds postulates something beyond what we can ever perceive, but does not postulate God. I agree, but my point is that one might as well postulate God, since this is not physics. As a physicist, I am opposed to the claim that either God or the Multi-verse is part of physics.

*Author:* If physics is about the physical world, then so is this theory. The problem here is not a God-bias but human-bias, e.g. that *our* consciousness creates quantum collapse, that *our* minds create the world, or that what we see is what is *because we see it*. Our senses are biological accidents - we see *a* world not *the* world, e.g. bees ultra-violet and bats sonar senses see a different world. Our technology sees x-rays and nature's eye sees light, but whether by technology or biology, each is only a "view" of the world. Logical positivism is the naive nineteenth century idea that *we see what is*. It masquerades today as an axiom of science, but really is not part of science at all. Neither solipsism, that the mind creates everything, nor physicalism, that only what we sense exists, have any solid foundation. To define reality as what we see or know is human-centric bias. Quantum theory directly challenges our existential self-centeredness because it mathematically references quantum waves we can't see. Its "measurement paradox" denies science in principle the chance to see quantum waves directly, as every observation gives only a "thing" in one place:

"*The full quantum wave function of an electron itself is not directly observable…*" [19] p240.

When Pauli and Born defined the quantum wave as an existence probability amplitude, physics ceased to be about anything physical at all:

"*For the first time in physics, we have an equation that allows us to describe the behaviour of objects in the universe with astounding accuracy, but for which one of the mathematical objects of the theory, the quantum field ψ, apparently does not correspond to any known physical quantity.*" [38] p89

If science knows the world by observing, and observing always collapses the wave function, we can't directly study quantum waves. Since in logical positivism only "*…what impinges on us directly is real.*" [39] p9, unobservable quantum states can't be real. Either quantum theory unscientifically references the unobserved, or as Penrose says, the "unreal" predicts the "real". Yet the experimental method itself references probabilities, which we cannot see or touch, and so also "don't exist". So by positivism, science is not scientific! Conversely, if quantum theory is indeed scientific then logical positivism is not a necessary condition for science. That *we* can't see quantum waves directly is emphatically not a reason to conclude that they don't exist. The nineteenth century doctrine of positivism rightly fought the superstitions of the middle ages, but it has had its day. Two centuries later, we need not fear a return to pre-science fallacies by merely *theorizing* that the physical world is incomplete. Certainly VR theory deposes physical realism, but it restores mathematical realism!

*Comment:* An entity (SuE) is in a collapsed quantum state (pure real state) just for a moment the "time moment" that it is on time line (X=Y=Z); the time of perception (measurement). The same is true for the "time moment" when it is in its pure virtual state $\Phi X = \Phi Y = \Phi Z$ . Hence, any entity is in its clear real or virtual state just for a "time moment" when it (the entity) is on its time line (extension of R or $\Phi$, see Fig. 5 and 6 for 2D and Fig. 8 for 3D). At all other (infinity) time the entity is in a sum of stochastic states with continuously varying probabilities between the previous and the following collapsed quantum states. So reality is neither real nor virtual.

*Author:* That is a reasonable conclusion by the evidence. The VR conjecture just raises another option - that uncollapsed quantum waves are the ongoing reality and "physical reality" the output, i.e. when programs overload the grid and restart. If the "pure real states" we see only ever exist for an instant, as quantum theory says, they are *events* not states. In this model, physical events are entity program reboots. If the physical world is built entirely from such events, it has no "substance" in itself.

*Comment:* Reality is that which causes ONLY IF interaction is independent from what is caused. If this is not the case (interaction depends on the result through holographic virtual interconnection) then reality is that which causes PLUS that which is caused. It is not like Plato's cave. It is like we see with one eye to our front and the other eye to our back. The front eye sees the real reality and the back one sees the virtual reality. We can not perceive both realities at once although they are both existing in the present universe. We have the appropriate glass(es) for the front eye and we are search for the glass(es) for the back one.

*Author:* This is a dualist position, so has the problems of dualism, e.g. if quantum waves cause physical reality which then causes quantum waves, which came first, to begin the interaction? Or if both exist in their own way, why is the quantum world so different? How can it do what can't be physically done? Why have one set of rules for the physical world and another for the quantum world? Non-dualism is simpler, and so by Occam's razor preferred. Moving our "reality" focus from the physical to the quantum world, lets the latter create the former as processing creates an output. All that is required is to abandon physicalism. In contrast, dualism needs two realities, as your glasses analogy illustrates. In science we register the physical world by our senses, or as you say, the "first eye". Technical instruments like the telescope are "glasses" to aid the eye of our senses. If there is also a "back eye", that sees another world, how does it arise apart from the brain? How does it register a non-physical world? What "glasses" would help a non-physical eye in its observations?

*Comment:* Your essay seems unique in its description of processing and programs. Some things could be argued but those would naturally go away as the model is refined with more specific detail on how it works. The concepts of local processing, how it is used up, and the rip are interesting, but seem to ask for more definition. I found most interesting the photon Planck program. Do you have any thoughts of defining pseudo-code for this program? A code definition would help tighten up and logically define the ideas described. Could it be written in a standard programming language or simulated with a standard computer?

*Author:* Yes, dynamic programs not static formulae are the way forward, as the failure of string theory illustrates. Physicists like static formulae but Nature's power needs recursive languages like Lisp to express recursive algorithms like Mandelbrot's [40]. The Planck program proposed is just a moving discrete rotation that gives a sine wave, and expanding it gives the spherical wave of Schrödinger's equation. After that, the formulas get too hard, but we can simulate complex problems with valid simplifying assumptions, as traffic simulations do. Currently, we are like the man looking for his keys under a lamp post, who when asked where he lost them said "*In the dark forest over there but the light is better here*", e.g. we assume a flat 3D Euclidian space when we know space curves; we assume triangular spin networks when even war gamers use hexagons not squares; we assume static links when even cell phone networks dynamically reorganize under load. Yet even simple programs that use a cubic lattice can give useful results [41].

*Comment:* You say: "*if the mind creates the body as in a dream, why can't I dream the body I want?*" [1] but the mind doesn't create the body but it does create an image of the body. I propose a simple answer given by Hoffman in his Interface Theory of Perception.

*Author:* The quote was arguing for exactly what you say. As originally a psychologist, I accept that the brain constructs *its* reality, rather than veridically reflecting what is out there. This doesn't mean our minds create all physical reality, as some falsely assume, but quantum mechanics implies just that:

*To the extent that it {a photon} forms part of what we call reality… we have to say that we ourselves have an undeniable part in shaping what we have always called the past.* [42] p67.

Only if all physical reality actually is virtual is the above sensible. When one observes, physics and psychology are different system views. To be consistent, one must pick an observer stand-point and stick to it, i.e. one cannot "mix-and-match" observer perspectives [43].

*Comment:* Your essay is very clearly expressed and nicely illustrated. I think I would have enjoyed it but due to my own opinion on the nature of reality I experienced discomforting cognitive dissonance through out. I wanted to immediately protest at some of the assumptions and arguments made.

*Author:* You well describe how many feel when they read this. I would feel the same had I not written it! Yet unsettling as it is, the question deserves consideration, because people are asking it.

*Comment:* What makes something real? Analogy ... Is the software more real than the screen display, the avatars that enact the game, and the visual experience of the player, or are they all real in their own way?

*Author:* Re what is reality, your example is telling. A computer game can be "real" to one so involved in it they see nothing else. It is a *local reality*, a world real from within but not from without. Yet by *identification*, an "unreal" virtual event can affect a "real" observer, e.g. in Japan, an assault court case arose because in an online game, a player lent his special sword to another avatar in the game, who then sold it on e-bay! So was the sword that was stolen real? Let us define reality as *what exists at the end of the causal line*, i.e. an uncaused existence, e.g. in The Matrix, the construct world was created by machines in a world where Neo is a body in a vat, so was unreal. In contrast, the physical world that the machines and Zion inhabit was uncaused by anything else, so was real. This model is NOT The Matrix because in it the physical is virtual, full stop. Physical objects are assumed to self-exist and so be real, e.g. as the sun's light shines on earth, its photons exist in and of themselves so are real. In contrast, my image in a mirror is only there when I look, i.e. its existence is caused, so it is unreal. We see our world as real because we see it as uncaused. Now quantum theory tells us that the physical world is the latter case, i.e. it only exists *when we look at it*, like the image in the mirror (except we also get the "reflections" of others). This model also argues the same - that the physical world is generated.

*Comment:* Do you count all observation and experience of the world through the senses as unreal? Where does that leave practical science and the scientific method?

*Author:* In an online multi-player game, pixel avatars cause virtual experiences that produce real impacts on us. The pixels are virtual but the observers are real. So that the physical world we experience is virtual doesn't necessarily make it a fake. Suppose one day the processing behind the virtual online world The Sims allowed some avatars to "think" of themselves. To practice science, they would only need information to test theories against, which the virtual reality could provide. If they found their world had a malleable space-time and a maximum speed, etc. they might conclude it was virtual, from how it behaved. Not only does science allow a virtual reality but a virtual reality could also allow science.

*Comment:* In a way you sidestep the question, by making a conjecture and saying if it is true, then reality is digital. It only considers an underlying reality, which leaves out a vast amount of "otherness" that could potentially also be regarded as reality.

*Author:* To conjecture the world is virtual (p1) then say it is digital (p11) would sidestep the question, but in between are many pages on why this is possible and how it could be so. What about the many other human things not covered? This model is about physics, not psychology or philosophy. If you wonder if it supports new age ideas of telepathy, psychokinesis, aliens, crop circles, etc, I don't think so. Why should a simulation let avatar programs change the operating system rules? If I was the system administrator I wouldn't allow that. Yet equally, it doesn't deny that possibility.

*Comment:* About discreteness I think there is nothing fundamental about it, we see discrete features on quantum reality just because we use tools based on classical logic to get a partial understanding of the quantum world.

*Author:* Even with quantum tools, Heisenberg's uncertainty principle still applies. Discreteness is fundamental to information processing, as information is defined as a choice from a set of options. If that set were infinite it could not be enumerated to make the choice, i.e. every processing output must be discrete, so that the physical world is discrete is built into the VR model. Whether the physical world is discrete or not is an open question, but evidence includes:

1. Planck limits on length, time, energy etc suggest measurement discreteness.
2. Heisenberg's uncertainty principle defines **h** as the discrete value.
3. Photon wave energy quantization suggests that wavelength changes are discrete
4. Non-discrete continuity creates infinities and paradoxes, e.g. Zeno.
5. A discrete world with no infinitely small has no infinitely large, hence space has a finite capacity and light has a maximum speed.
6. In calculus infinitesimals can "tend to zero", i.e. a small discrete value can in one step become zero. Calculus then works because the physical world really is discrete.
7. Spin networks, loop quantum gravity and quantum simulations successfully assume discreteness.
8. Cosmological models suggest fundamental upper bounds on the information processing of space, which then imply lower bounds of discreteness.

If the physical world is not discrete or if it is ever infinitely large, this model is immediately falsified.

*Comment:* What a pleasure to read your essay! Are you making any new virtual reality games? Please let me know.

*Author:* Why bother, as we already have one that has run successfully for billions of years?

*Comment:* Are you sure it is just one running? Could it be several running simultaneously?

*Author:* How can several run simultaneously given each observer only sees one? Or if many observers "see" different views, isn't that the same as several? By Occam's razor, if one world suffices for us scientists, why postulate more? Everett invented a fantastic multiverse machine to exorcize the quantum ghost, but just accepting quantum reality is the simpler option. If our universe is simulated on the inside of a hyper-space bubble expanding into a larger bulk, there could be many other such bubbles. A system that creates one simulation could indeed create others. Who knows?

*Comment:* This work is deeply meaningful, whether it turns out to be 'spot on' or not. Your proposal that we need a theory that is 'background independent' and 'foreground independent' may also be historic. I think the section on the 'Planck program' is somewhat reminiscent of papers and talks by B.G. Sidharth, but by and large what you are presenting is highly original. I agree with almost everything you say, except the very last sentence. I think there is a bias toward the idea that computers are digital, by nature, but a universal quantum computer that lives 'outside the universe?' Well maybe.

*Author:* Yes you are right, who knows what the "other" is, as we only know the world we see and register, as positivists rightly state. The argument was that if the physical world we see is created by processing, then it must be digital, by definition. It does not speculate on any containing reality.

*Comment:* If one assumes that the 'processor' is 'otherworldly', as you do, I don't know why one wouldn't assume the existence of 'perfect' components used to build the processor, and, in that case, there is no reason that is obvious to me that the processing could not be analogue, and not digital. If each 'node' on your 'grid' is an analogue processor, suitably connected to other nodes, there is no evident reason, other than current technology and economics biases, to assume digital. So your conclusion, "The argument was that if the physical world we see is created by processing, then it must be digital, by definition" seems unwarranted.

*Author:* I used the word "other" not "otherworldly". It is not my word but Fredkin's, the computer scientist who began the VR idea twenty years ago [45]. To say this model *assumes* processing in another world misrepresents it. It just asks a question of the physical world, namely, is it a virtual reality, created by processing? Science, as a way to ask questions of the world, allows the question. It then assumes it is true, as a hypothesis, to check its implications against the world we see. One of these is that a virtual reality needs a containing reality, as a system cannot output itself, just as a printer cannot print itself out. So that there is an "other" is a *conclusion* not an *assumption* of this conjecture, i.e. it only arises if it is validated. So it doesn't specify that "other" except that it processes. Speculations about it, including yours, that don't reference the world we know, are scientifically idle, e.g. that:

1. Our universe could be "saved" and "restored" [46].

2. Our virtual reality could be created by another [38]

3. Every quantum event creates a new universe [35]

This is science fiction not science. In contrast, asking if the physical world is created by processing is science, as it has implications for how the world behaves. Yes the containing reality could be analogue but this model applies to its information processing output. Shannon and Weaver define information as the number of options of a choice [11]. If that number is infinite, the options cannot be enumerated, so the choosing cannot occur. So an information output, even by an analogue processor, is always a finite total of discrete bits. In no case do our processors output digital values that are infinite. In this respect a qubit is just as digital as a bit, as its choices are equally finite. So the conclusion that if the physical world is a processing output it must be digital remains.

*Comment:* "Nor is this solipsism, that the physical world is just a dream, which Dr Johnson is said to have refuted by stubbing his toe on a stone, saying "I disprove it thus". Is it your position that George Berkeley believed the world to be a dream?

*Author:* Solipsism is that the human mind creates the physical world as a dream. Bishop Berkeley's claim that the senses create the world is a basis of perception psychology today, e.g. optical illusions show that the brain indeed "manufactures reality". This "*esse ist percipi*[5]" thesis, that things have no existence apart from our perceiving them, applies to each of us individually. It does not necessarily contradict realism, that a world exists apart from us. I don't know if Berkeley recognized this but gather he was more sophisticated than his critics made out. This paper opposes human-centrism, e.g. that the universe was made for us, that it needs us for quantum collapse or was in a superposed state for billions of years until we came along to "observe" it into existence. Physicalism, that the physical world exists in and of itself, is also human-centric [2] (p6), as it defines reality as what we register. If science tells us anything, from Copernicus to Darwin to Einstein, it is that we are *not* the centre of the universe. To take *our* reality as *the* reality is egocentrism all over again. In this model, the physical world is virtual but there is a real world out there, apart from us, i.e. it is not solipsism or nihilism. It has a real world, but just not the world we see. When we observe an electron say, we see a "particle" in one place, but before and after that, in this model, it was just spreading waves of processing, i.e. non-local. Our observing caused the system event, specifically a program reboot, we call a physical particle. So in this model, quantum mathematics describes what is actually there - the distributed processing whose output is our physical reality.

---

[5] "Existence is perception".

*Comment:* "Randomness. If every physical event is predicted by others, a random quantum event is an impossible "uncaused cause", but a processor creating a virtual world can be its cause." I assume you are using 'random' to have a technical meaning. In other words, not meaninglessness.

*Author:* A random event is one that no preceding physical event combination predicts, i.e. no physical world "story" leads up to a random event. Such events should not arise in a causal self-contained physical world, but they do in ours. This doesn't necessarily imply "free-will", but it does let it off the hook of determinism. While randomness has no meaning to individuals, choice *from outside a system is necessary for it to evolve*, whether it is matter or animals evolving. An environment can only induce an evolution given free choices to generate a variety that it can "naturally" select from.

*Comment:* Your essay is well written, well explained and well presented but you confound computing and the universal dynamic. This is just a play of computing. The realism never will be other than this pure objectivity of uniqueness and its entropy. You can compute emergent universes on the 2D picture after all, but frankly for the convergences in 3D? Let's be serious a little. If you can create a flower with your computer, tell me it ....

*Author:* Everything depends on where you view from. I can indeed create a flower using my computer, but it will be a digital one (2D or 3D). To me it is not a "real" flower, but an avatar who sees only digital "things" might consider it as real as the rest of his virtual world. Who then is to say that the "real" flower I see outside my window is not also created thus? What proves that my world is objective? Its a matter of perspective - a virtual world unreal from the outside can be real from within. The movie 'The Matrix' made this point brilliantly, but cunningly kept the ultimate reality physical to keep the fan base - Neo exits the matrix into a physical world. The VR conjecture in contrast "thinks the unthinkable", that physicality itself is virtual although it seems real to us. The paper doesn't talk about computing but *processing*, whose definition implies no physical base. So classical computing needs a physical base but quantum computing doesn't.

*Comment:* You are right about virtual reality, this the nature of the relativistic view of the universe and hence it is digital. The absolute view of the universe is singularity and it is analogue. Singularities in physics are like unexplainable points in space-time. In the relativistic view (relativity theories) of the universe there seems to be multiple singularities in multiple black holes. But upon further understanding the scientific world will realize that all these black holes are indeed connected and there is only one singularity at the heart of them all. I hope that I have conveyed the truth in simple enough words to touch the scientific hearts. As a father of three boys on this planet earth, I could now being fully aware of myself say that I created the universe for them.

Duality is to think that I am not the universal I.

Singularity is to know that I am the universal I.

But what I see as it is in the world is far from what I want this world to be. We are fighting in the name of the very foundations we were supposed to be united and this will lead us to extinction. It is time for us in this duality to realize the singularity of love and live in peace.

*Author:* I don't think singularity means what you think it means. In physics, it is an equation infinity, which is not unusual. This model has no singularities. Its big bang was *one* photon that by a cascading inflation process tore the grid apart to create a universe of oscillations on a 3D surface. The ripping stopped as the expanding surface created, our space, diluted the waves. So the universe never did all exist at a point. Nor are black holes singularities - just the grid processing at full capacity. Or if you are saying that the universe is a singular thing, then physical realists who see it as a *closed system* would agree, e.g. as the physical atoms of our bodies were once stardust, so they will again return to the physical "one". Or if you mean that "I" am "You", and we are both an absolute universal "I", then why discuss? Why does "I" talk to "I"? How is there a universal "I" with by definition no "you" outside it? Do you mean a universal "We"?

This theory is not that all is one, or that the mind creates the universe, or that the universe is holistic, or that nothing really exists at all. It is that all physical reality is in essence virtual. The physical world is then a bubbling flux because, like a TV screen picture, it must be continuously refreshed to exist. This includes the phenomenal world of sensations, actions, causes and results, i.e. not just the physical body, but also the brain and the sensations and ideas it generates - including those of love, harmony, self, other selves or a universal I. So everything observed, done, said or thought is virtual, i.e. a quantum collapse. To want to change the physical world to be of harmony and love is to want a better virtuality, a better looking screen output, while ignoring the program behind it. Wanting to "fix" the world assumes it is broken, but at the quantum level it is working perfectly, as a karmic output system. If each output is an input, etc, how can anything change, as this flawless machine grinds causality by trying every option? Perhaps the way out is in, i.e. consciousness. Information processing needs a source, here the grid, so our *sense of being an observer,* if that is what you mean by "I", may be valid. Even in a virtuality, the choices made are real. So if we have choice, we can rewrite the program at source by changing our *being*. A program can't change the operating system, nor other programs, but it could alter itself. Spirituality is then not about rearranging inputs and outputs, as in politics and power, but about changing the program source, i.e. not about changing the physical world but about changing ourselves.

## References

[1]    Whitworth, B., "The emergence of the physical world from information processing," *Quantum Biosystems*, vol. 2, no. 1, pp. 221-249, latest version at http://brianwhitworth com/BW-VRT1 pdf. 2010.

[2]    B. Whitworth, "Simulating space and time," *Prespacetime Journal*, vol. 1, no. 2, pp. 218-243, latest version at http://brianwhitworth.com/BW-VRT2.pdf, 2010.

[3]    B. Whitworth, "Ch. 3: The light of existence," in *The Physical World as a Virtual Reality*, http://brianwhitworth.com/BW-VRT3.pdf, 2012.

[4]    B. Whitworth, "Ch. 4: The matter glitch," in *The Physical World as a Virtual Reality*, http://brianwhitworth.com/BW-VRT4.pdf (in preparation), 2012.

[5]    K. Zuse, *Calculating Space*. Cambridge Mass.: MIT, 1969.

[6]    S. Lloyd, "Universe as Quantum Computer," *arXiv:quant-ph/9912088v1*, vol. 17, 1999.

[7]    G. McCabe, "Universe creation on a computer," *Stud.Hist.Philos.Mod.Phys.36:591-625*, 2005.

[8]    L. Smolin, *Three Roads to Quantum Gravity*. New York: Basic Books, 2001.

[9]    C. E. Shannon and W. Weaver, *The Mathematical Theory of Communication*. Urbana: University of Illinois Press, 1949.

[10]   B. D'Espagnat, - *On physics and philosophy*. Princeton, NJ.: Princeton University Press, 2006.

[11]   P. Zizzi, "Emergent Consciousness; From the Early Universe to Our Mind, arXiv: gr-qc/0007006," *NeuroQuantology*, vol. 3, pp. 295-311, 2003.

[12]   F. Wilczek, *The Lightness of Being: Mass, Ether and the Unification of forces*. New York: Basic Books, 2008.

[13]   B. Whitworth, "Ch. 2: Simulating Space and Time," in *The Physical World as a Virtual Reality*, http://brianwhitworth.com/BW-VRT2.pdf, 2012.

[14]   R. P. Feynman, R. B. Leighton, and M. Sands, *The Feynman Lectures on Physics*. Reading, Ma.: Addison-Wesley, 1977.

[15]   J. Case, D. S. Rajan, and A. M. Shende, "Lattice computers for approximating euclidean space," *Journal of the ACM*, vol. 48, no. 1, pp. 110-144, 2001.

[16]   Edwin Abbott, "Flatland: a romance of many dimensions," *Project Gutenberg*, 1884. [Online]. Available: http://www.gutenberg.org/etext/201. [Accessed: 22-Feb-2010].

[17]   G. P. Collins, "Computing with quantum knots," *Scientific American*, vol. April, pp. 56-63, 2006.

[18]   S. W. Hawking and J. B. Hartle, "The basis for quantum cosmology and Euclidean quantum gravity," *Phys. Rev.*, vol. 28, no. 2960, 1983.

[19]   L. M. Lederman and C. T. Hill, *Symmetry and the beautiful universe*. New York: Prometheus Books, 2004.

[20]   R. B. Laughlin, *A Different Universe: Reinventing physics from the bottom down*. New York: Basic Books, 2005.

[21]   B. Whitworth and A. Whitworth, "The Social Environment Model: Small Heroes and the Evolution of Human Society," *First Monday, Volume 15, Number 11 - 1 November 2010*, vol. 15, no. 11, p. http://firstmonday.org/htbin/cgiwrap/bin/ojs/index.php/fm/article/view/3173/2647, Nov. 2010.

[22]   A. Guth, *The Inflationary Universe: The Quest for a New Theory of Cosmic Origins*. Perseus Books, 1998.

[23]   W. M. Itano, D. J. Heinzen, J. J. Bollinger, and D. J. Wineand, "Quantum Zeno effect," *Physics Review A 41*, vol. 5, no. 1, pp. 2295–2300, 1990.

[24]   J. Audretsch, *Entangled World: The fascination of quantum information and computation*. Verlag: Wiley, 2004.

[25]   J. Barbour, *The End of Time: The next revolution in physics*. Oxford: Oxford University Press, 1999.

[26]   J. B. Hartle, "The Physics of 'Now'," *Am.J.Phys.*, vol. 73, pp. 101-109, avail at http://arxiv.org/abs/gr-qc/0403001, 2005.

[27]   L. Smolin, *The Trouble with Physics*. New York: Houghton Mifflin Company, 2006.

[28]   R. Penrose, *Shadows of the Mind*. Oxford: Oxford University Press, 1994.

[29]   J. Young, *Imagining Our World as a Virtual Reality, http://chronicle.com/article/Audio-Imagining-Our-World-as/63403/*. .

[30]   B. D'Espagnat, "The quantum theory and reality," *Scientific American*, vol. 241, no. 5, pp. 158-182, 1979.

[31]   A. Aspect, P. Grangier, and G. Roger, "Experimental Realization of Einstein-Podolsky-Rosen-Bohm Gedankenexperiment: A New Violation of Bell's Inequalities," *Physical Review Letters*, vol. 49, no. 2, pp. 91-94, 1982.

[32]   B. Whitworth, "Politeness as a Social Software Requirement," *Int. J. of Virtual Communities and Social Networking*, vol. 1, no. 2, pp. 65-84, http://brianwhitworth.com/polite3.pdf, 2009.

[33]   B. Whitworth, "A Comparison of Human and Computer Information Processing," in *Encyclopedia of Multimedia Technology and Networking*, M. Pagani, Ed. Hershey PA: Information Science Reference, 2009, pp. 230-239, http://brianwhitworth.com/braincomputer.pdf.

[34]   J. Conway and S. Koch, "The free will theorem," *Found. Phys.*, vol. 36, no. 10, p. arXiv:quant-ph/0604079v1, 2006.

[35]   H. Everett, "'Relative state' formulation of quantum mechanics," *Rev. of Mod. Phys.*, vol. 29, pp. 454-462, 1957.

[36]   M. Tegmark and J. A. Wheeler, "100 Years of the Quantum," *Scientific American*, no. Feb, pp. p68-75, 2001.

[37]   E. H. Walker, *The Physics of Consciousness*. New York: Perseus Publishing, 2000.

[38]   N. Bostrom, "Are you Living in a Computer Simulation?," *Philosophical Quarterly*, vol. 53, no. 211, pp. 243-255, 2002.

[39]   R. Oerter, *The Theory of Almost Everything*. London: Plume, Penguin, 2006.

[40]   N. D. Mermin, "Whats bad about this habit?," *Physics Today*, vol. May, 2009.

[41]   J. Dickau, "How Can Complexity Arise from Minimal Spaces and Systems?," *Quantum Biosystems*, vol. 1, pp. 31-43, 2007.

[42]   B. Maier, *Reality is Virtual*. Shoreham, New York: Infinity Publishing, 2010.

[43]   P. Davies and J. R. Brown, *The Ghost in the Atom*. Cambridge: Cambridge University Press, 1999.

[44]   B. Whitworth, "The social requirements of technical systems," in *Handbook of Research on Socio-Technical Design and Social Networking Systems*, B. Whitworth and A. De Moor, Eds. Hershey, PA: IGI, 2009, http://brianwhitworth.com/STS/STS-chapter1.pdf.

[45]   E. Fredkin, "Digital Mechanics," *Physica D*, pp. 254-270, 1990.

[46]   J. Schmidhuber, "A Computer Scientist's View of Life, the Universe and Everything," in *Foundations of Computer Science: Potential-Theory-Cognition Lecture Notes in Computer Science*, C. Freksa, Ed. Springer, 1997, pp. 201-208.

## Notes

---

[1] In the analogy, people are tied up in a dark cave with their backs to its exit. Looking at the cave wall, they see their shadows, created by sunlight from the outside, and take the shadows as their reality.

[2] For example, Fredkin, Lloyd, Tegmark, Campbell, Svozil and Wolfram.

[3] For example: "*Imagine the quantum computation embedded in space and time.*" [6] p172.

[4] Biological properties can evolve by autopoietic bootstrapping, but existence itself cannot arise the same way. No amount of "emergence" by the interaction of existing parts can create something from nothing.

[5] McCabe argues that the physical world can't be simulated as follows: "*All our digital simulations need an interpretive context to define what represents what. All these contexts derive from the physical world. Hence the physical world cannot also be the output of such a simulation*" [7]. The logic is correct for static data which needs a viewer context, but not for the dynamic processing of this model, which doesn't. See [2] p5 for details.

[6] Information $I = Log_2(N)$ for $N$ *finite* options in a choice.

[7] By the quantum *indistinguishability* principle, it is impossible to mark any electron apart from another.

[8] Berkely developed solipsism from Platonic idealism.

[9] So quantum collapse is not just if we observe the world, but if anything does, i.e. any quantum interaction.

[10] In science fiction terms, if aliens are creating a virtual reality from outside it, *we are the aliens*.

[11] In physicalism, an object like a rock is not conscious. In this model it is the opposite, as *everything* has consciousness in it, including a rock. What distinguishes us from a rock is self-awareness, not consciousness.

[12] Instantiation of an entity class is an object orientated systems (OOS) method by which identical information "objects" dynamically inherit code from a single program blueprint, e.g. screen buttons instantiating the same class run the same code, so all look the same. A class dynamically feeds code to its many instances on request.

[13] Actually, a node planar channel overloads, where a channel is a plane through a node. See [13] p 23

[14] This doesn't repeat the overload because nodes first share and then process. See [13] p16 for the details.

[15] We call it random because it depends upon no previous physical event.

[16] In relativity, light doesn't travel in any self-evident "straight line".

[17] As Feynman states: "Whatever is not explicitly forbidden must happen". Gell-Mann called it the quantum totalitarian principle.

[18] Only at a particular processing cycle does one space "point" necessarily map to one grid node.

[19] Planck length of $10^{-33}$ meter is the pixel resolution and Planck time gives a refresh rate of $10^{43}$ Hertz.

[20] These are:

   a) If a tortoise running from a hare sequentially occupies infinite points of space, how can the hare catch it? Every time it moves to where it was, the tortoise has moved on.

   b) Or, if space-time is not infinitely divisible, there must be an instant when the arrow from a bow is in a fixed unmoving position. How can a set of static instants beget continuous movement?

To accept either one is to trip up on the other, i.e. without an infinity of points there is no continuous movement. Or if there is an infinity of points, the hare can never catch the tortoise.

[21] Like Mr. A. Square of Abbot's Flatland, we struggle to imagine a dimension beyond ours [14].

[22] One can imagine hyperspace as space but with every point a tiny sphere.

[23] If N is the grid granularity, the local Planck event angle is 360°/N.

[24] For example, "*In quantum mechanics there **really are** complex numbers, and the wave function **really is** a complex-valued function of space-time.*" [19] p346

[25] For example, "*… we accept as nonexistent the medium that moves when waves of quantum mechanics propagate.*" [20] p56.

[26] Some of course believe this is possible, see our review of the credit meltdown [21].

[27] Space is *isotropic*, i.e. all directions are equivalent, so vibrations in it <u>cannot</u> generate positive-negative valences.

[28] The Michelson–Morley experiment denied the idea of a physical ether, but a non-physical ether has never been contradicted. Indeed that space is filled with an invisible "quintessence" has waged a vigorous comeback.

[29] In this model, quantum mathematics is not just a "convenient fiction". Electro-magnetism really does oscillate into an "imaginary" dimension outside "real" space, as complex number theory says.

[30] The physical analogy is to help understanding only, and should not be taken literally.

[31] In this model, a pixel is a quantum state.

[32] The principle of conservation of processing, that every entity program instruction is always allocated to run on the grid only once, is proposed to underlie all our partial conservation laws, of mass, energy, charge, etc.

[33] In current physics, $E.\lambda = h$, with E wave energy, $\lambda$ wavelength and h Planck's constant. In the model, the processing rate across a photon's wavelength (Planck's constant) is its processing rate per node length (wave energy) times by the number of nodes involved (wavelength), i.e. the model fits the formula.

[34] This bears on what physics calls the "hierarchy problem".

[35] In this model, the inner surface of the grid "hole" began our space and the "first photon" began our universe.

[36] In this model, a packet transferred once per node cycle travels at the speed of light. In inflation, each transfer <u>immediately</u> "rips" the receiving node apart, before it can process any cycle. If each cycle "tick" is constituted of many value "clicks", the chain reaction occurs at the click rate not the tick rate, i.e. <u>much</u> faster than the speed of light tick rate, e.g. estimates of a $10^{50}$ increase in $10^{-33}$ seconds.

[37] Extreme light has an extreme photon in every polarization plane of its movement axis. An extreme photon is a maximum frequency photon, i.e. with a two Planck length wavelength.

[38] As in this model grid nodes are not exactly synchronous, there will be a slight mass.

[39] Including the equations of quantum mechanics.

[40] In what Barbour calls a "time capsule", where all states past and future exist in a "timeless" universe [25].

[41] In a "*Physics of Now*" [26] p101, each "state" is equivalent to another choice.

[42] The basic grid existence operation, a transverse rotation "add one", is asymmetric with respect to the hyper-surface, i.e. one can also "subtract one". Only if existence is processing can it run in reverse, as anti-matter.

[43] The first value set can be "out" or "in" with respect to a sphere surface. This rotation direction choice defined our universe as being of matter or anti-matter. There is no need to wonder "*Where did all the antimatter go?*", as current physics does, because it didn't go anywhere, as it never was. It only occurs now in special cases, as a system option, but the asymmetry occurred at the beginning.

[44] This model preserves the causality and unpredictability that the following paradoxes demand:

  a) In the *grandfather paradox* a man travels back in time to kill his grandfather, so he could not be borne, so he could not kill him. One can have causality or travel back in time but not both.

  b) In the *marmite paradox* I see forward in time that I will have marmite on toast for breakfast but then choose not to, so didn't rightly see forward in time. One can have choice or predictability but not both.

Time as grid processing is causal, and events as node reboots are unpredictable, i.e. there is no time travel.

[45] In the apocryphal story, a scientist lecturing that the universe needed nothing outside itself was challenged by a little old lady, who said it sat on the back of a giant turtle. He laughed, and asked her what the turtle was standing on, but got the reply "*Sonny, it's turtles all the way down*".

[46] In an electron-type channel overload (Figure 7), the mass $m_P$ is by definition the number of Planck program instructions, or values set. Also by definition, the energy of a transverse rotation is the rate $m_P$ values are set over a node distance, i.e. $m_P.L_P/T_P$, for Planck length $L_P$ and Planck time $T_P$. As the speed of light $c=L_P/T_P$, this is $m_P.c$. The photon's transverse processing also moves forward at speed of light, $c=L_P/T_P$, so the final energy is $m_P.c^2$. So for any transverse wave rebooting in a node, $E = mc^2$. If all mass arises the same way, the equation is general. This derivation is not a proof - Einstein did that from how the physical world works - but shows that Einstein's equation is compatible with this model.

[47] Or bosons and fermions.

[48] Colliding protons can give new matter particles but still remain intact

[49] Relativity does finds the space-time combination absolute. In this model, finite grid nodes carry out both existence processing and movement transfer. In contrast, an objective space-time continuum has no common basis for space and time. If both are just "dimensions", why has time in the Lorentz equation a different sign?

[50] In the story of the stranger, a father and son boarded a train and at their meals another man congenially joined in and partook. Eventually the father complained to the son "Your friend eats a lot!", when the son replied "I thought he was your friend?" Upon being exposed, the imposter disappeared and was not seen again. The story shows how ill-founded assumptions install themselves, as everyone thinks the other has checked them.

[51] As Penrose says: "*How, indeed, can real objects be constituted from unreal components?*" [28] p313

[52] The nineteenth century doctrine of positivism has become a religious canon. Yet science is a way to study reality, not a belief about it. It is a way to ask questions, not a set of answers. See http://researchroadmap.org

[53] In Douglas Adams "The Hitchhiker's Guide to the Galaxy", the computer Deep Thought after millennia of calculations found that the answer to life, the universe and everything to be 42. It was, of course, a joke, but this model's proposal, that the physical world is a calculation, is not a joke.